\documentclass[a4paper,11pt]{article}
\pdfoutput=1 

\usepackage{pub} 

\usepackage[T1]{fontenc} 
\usepackage[italian,english]{babel}
\usepackage{hyperref}
\usepackage{ifpdf}
\usepackage{subfigure}
\usepackage{tikz}
\usetikzlibrary{arrows,shapes}
\usetikzlibrary{trees}
\usetikzlibrary{matrix,arrows} 				
\usetikzlibrary{positioning}				
\usetikzlibrary{calc,through}				
\usetikzlibrary{decorations.pathreplacing}  
\usepackage{pgffor}							

\usetikzlibrary{decorations.pathmorphing}	
\usetikzlibrary{decorations.markings}
\tikzset{
	vector/.style={decorate, decoration={snake}, draw},
	fermion/.style={draw=black, postaction={decorate}}, 
	scalar/.style={dashed,draw=black, postaction={decorate}}}
\tikzstyle{block} = [draw, rectangle, 
minimum height=3em, minimum width=6em]

\usepackage{amssymb}
\usepackage{amsfonts}
\usepackage{epsf}
\usepackage{rotating}
\usepackage{graphicx}
\usepackage{amsmath}
\usepackage{fancyhdr}
\usepackage{lineno}
\usepackage{babel}
\usepackage{graphics}
\usepackage{pstricks}
\usepackage{color}
\usepackage{multirow}

\newcommand{\bea}{\begin{equation}}
\newcommand{\eea}{\end{equation}}
\newcommand{\be}{\begin{eqnarray}}
\newcommand{\ee}{\end{eqnarray}}

\def\hbar#1{\backslash\hspace{-2mm}#1}

\catcode`\@=11
\def\lsim{\mathrel{\mathpalette\@versim<}}
\def\gsim{\mathrel{\mathpalette\@versim>}}
\def\@versim#1#2{\vcenter{\offinterlineskip
\ialign{$\m@th#1\hfil##\hfil$\crcr#2\crcr\sim\crcr } }}
\catcode`\@=12

\def\2tvec#1#2{
\left(
\begin{array}{c}
#1  \\
#2  \\
\end{array}
\right)}

\def\mat2#1#2#3#4{
\left(
\begin{array}{cc}
#1 & #2 \\
#3 & #4 \\
\end{array}
\right) }

\def\Mat3#1#2#3#4#5#6#7#8#9{
\left(
\begin{array}{ccc}
#1 & #2 & #3 \\
#4 & #5 & #6 \\
#7 & #8 & #9 \\
\end{array}
\right) }

\def\3tvec#1#2#3{
\left( 
\begin{array}{c}
#1  \\
#2  \\
#3  \\
\end{array}
\right)}

\def\to{\rightarrow}

\def\ntrli{\tilde{\chi}^0_i}
\def\ntrlj{\tilde{\chi}^0_j}

\newcommand{\ntrl}[1]{\tilde{\chi}^0_#1}
\newcommand{\chpm}[1]{\tilde{\chi}^\pm_#1}
\newcommand{\chmp}[1]{\tilde{\chi}^\mp_#1}

\def\hbar#1{\backslash\hspace{-2mm}#1}

\numberwithin{equation}{section}

\title{Long-lived triplinos and displaced lepton signals at the LHC}
\author[a]{Asl{\i} Sabanc{\i} Keceli}
\author[b]{Priyotosh Bandyopadhyay}
\author[c]{Katri Huitu}

\affiliation[a,c]{Helsinki Institute of Physics and Department of Physics, FI-00014 University of Helsinki, Finland}
\affiliation[b]{Indian Institute of Technology Hyderabad, Kandi,  Sangareddy-502287, Telengana, India}

\emailAdd{bpriyo@iith.ac.in} 
\emailAdd{katri.huitu@helsinki.fi}
\emailAdd{sabanciasli@gmail.com}
\preprint{IITH-PH-0004/18
	
	\hspace*{11.27cm}HIP-PH/2016/09/01}
\abstract{We explore the possibility of having superpartners of triplet Higgs bosons, named as 'triplinos'. They form a part of light neutralinos and charginos in a $Y=0$ extended supersymmetric Standard Model. For this model such electroweakinos do not have direct couplings to the Standard Model fermions. On top of that, due to very compressed 
	spectrum for lighter neutralinos and charginos, their decay products coming from three body decays are very soft and thus can evade  the current collider bounds. 
	These decays are particularly interesting since they give rise to displaced leptonic signatures.	We categorise the parameter space, while exploring different displaced decay possibilities.  A PYTHIA based simulation has been performed to find out the displaced 
	charged lepton, jet and $b$-jet final states at the  LHC with center of mass energy of 14 TeV. }  


\begin{document}
\maketitle
\flushbottom
\section{Introduction}
The discovery of the Higgs boson \cite{Higgsd1,Higgsd2} was the last piece in the Standard Model (SM), opening a new era in the particle physics. However, the LHC experiments have not ruled out the possibility of 
other scalars in the electroweak symmetry breaking (EWSB). The extension of the scalar sector in the context of supersymmetry is motivated by various reasons. Introduction of supersymmetry can solve the 
hierarchy problem, and along with conserved $R$-parity, it can give rise to a stable dark matter candidate. The discovery of a $\sim 125$ GeV Higgs boson in the minimal supersymmetric extension of the 
Standard Model (MSSM), demands the supersymmetric (SUSY) mass scale either to be very heavy or it requires large mass splitting between the superpartners of the top quarks \cite{carena}. This brings back the 
fine-tuning problem. An extension of the Higgs sector provides a solution to the problem with extra tree-level and loop-level contributions to the Higgs bosons mass. Thus  SUSY mass scale around TeV is still 
allowed \cite{PBAS}. Various such extensions include MSSM with a singlet \cite{NMSSM}, $Y=0$ triplet  \cite{epqr1,epqr2,PBAS,PBAS2, PBAS3}, $Y=\pm 1$ triplets \cite{agashe}, and the supersymmetric 
version \cite{quiros} of the Georgi-Machacek model \cite{GM}. Also supersymmetric extensions with both singlets and triplets have been studied extensively \cite{agashe, TNSSM1, TNSSMch}. In this article we 
focus on the extension of MSSM with $Y=0$ triplets \cite{epqr1, epqr2, PBAS, PBAS2, PBAS3}.

In particular, our focus is on the phenomenology of the electroweak gaugino and Higgsino sectors of the model. More precisely, the superpartners of the additional triplet scalars will mix with the standard MSSM 
neutralinos and charginos in the spectrum. In the gauge basis we call them 'triplinos'. One basic difference is that unlike gauginos and Higgsinos, they do not directly couple to the SM fermions and consequently to 
their superpartners. This feature affects the indirect bounds on the parameter space \cite{PBAS2}.  In this article we explore the production and decays of such triplinos at the LHC. For the neutral parts, the coupling to 
fermions and $Z$ boson comes via mixing to the $SU(2)$ doublets and hypercharged particles, which makes the phenomenology very interesting as we expect displaced decays of such triplinos (charginos and neutralinos). 
Occasionally the decay products of triplinos are so soft that they will be missed at the detectors and will give rise to disappearing charged tracks for triplet-like charginos. Such disappearing charged tracks have been 
investigated for some SUSY models at 8 and 13 TeV at the LHC experiments \cite{disch}. Here we investigate such scenarios for this model by segregating relative gauge hierarchy between the  lightest supersymmetric particle (LSP) and next to LSP (NLSP).

The article is organised as follows. In Section ~\ref{model} we give a brief introduction to the model along with the electroweak gaugino sectors. In Section~\ref{[param]} we find out the parameter spaces in the model consistent with the Higgs data  and the different kinds of NLSP-LSP scenarios. The collider simulation and related phenomenology are discussed in Section~\ref{pheno}. Finally we conclude in Section~\ref{concl}.

\section{Model}\label{model}
Triplet extended supersymmetric standard model (TESSM) contains a triplet chiral superfield $\hat T$ with zero hypercharge ($Y=0$) in addition to the MSSM fields. The scalar part $T$ can be represented as  a 2x2 matrix
\be
T=\left(\begin{array}{cc}\frac{1}{\sqrt{2}} T^0 & T^+ \\T^- & -\frac{1}{\sqrt{2}}T^0\end{array}\right),
\ee
where  $T^0$ is a complex neutral field, while  $T_1^-$ and $T_2^+$ are the charged scalar fields. Note that  $(T_1^-)^*\neq -T_2^+$. 

The renormalizable superpontential of TESSM includes only two extra terms as compared to MSSM, since the cubic triplet term is zero, and is given by
\be
W_{\rm TESSM}=\mu_T {\rm Tr}(\hat T \hat T) +\mu_D \hat H_d\!\cdot\! \hat H_u + \lambda \hat H_d\!\cdot\! \hat T \hat H_u + y_t \hat U \hat H_u\!\cdot\! \hat Q - y_b \hat D \hat H_d\!\cdot\! \hat Q- y_\tau \hat E \hat H_d\!\cdot\! \hat L\ ,
\label{SP}\nonumber
\ee
where other than the third generation Yukawa couplings are not included. Here "$\cdot$" represents a contraction with the antisymmetric  $\epsilon_{ij}$, with $\epsilon_{12}=-1$, and a hatted letter denotes the corresponding superfield.  $\mu_D$ is the usual mixing parameter of the two Higgs doublets and $\mu_T$ is the mass parameter of the triplet. Notice that while the triplet field $\hat T$  couples to the two Higgs doublets by a dimensionless coupling $\lambda$, the triplet-SM fermion couplings are absent due to the lack of right-handed lepton doublet in the theory. 

The soft SUSY breaking potential of the Higgs sector $V_S$ can be written by using the convention of the superpotential as
\be
V_S&= &m_{H_d}^2 |H_d|^2 + m_{H_u}^2 |H_u|^2 + m_{T}^2 {\rm Tr} (T^\dag T)
+[ B_D \mu_D H_d \cdot H_u + B_T \mu_T {\rm Tr} (T T)\\\nonumber
&&+A_{\lambda} \lambda H_d \cdot T H_u + y_t A_t \tilde{t}^*_R H_u\!\cdot\! \tilde{Q}_L- y_b A_b \tilde{b}^*_R H_d\!\cdot\! \tilde{Q}_L +{\rm h.c.} ] .\
\ee
Here $B_D$ and $B_T$ and $A_{j}$ ($j=\lambda,t,b$) are  the soft bilinear and the soft trilinear parameters respectively, while $m_i$ ($i=H_d,H_u,T$) represent the soft SUSY breaking masses.  Throughout this paper the parameters as well as the vacuum expectation values of the neutral Higgs fields (VEVs) are chosen to be real so that  Higgs sector does not introduce any CP violation. The EWSB is realised when the neutral component of the fields acquire non-zero VEVs, denoted by
\be
\langle H_u^0 \rangle = \frac{v_u}{\sqrt{2}},\;\; \langle H_d^0 \rangle = \frac{v_d}{\sqrt{2}},\;\; \langle T^0 \rangle = \frac{v_T}{\sqrt{2}}, 
\ee
and $\tan{\beta}={v_u}/{v_d}$. In this case, the $W$ boson mass expression is altered by the  triplet VEV as $m_W^2=g_2^2(v^2+4v_T^2)/4$, given $v^2=v_u^2+v_d^2$, whereas the $Z$ boson mass expression 
remains unaffected. This non-zero triplet contribution to $W$ mass leads to a deviation in the tree-level $\rho$ parameter expression, $\rho=1+4v_T^2/v^2$ and  the electroweak precision tests of the $\rho$ parameter 
\cite{Olive:2016xmw} impose  quite stringent constraint on triplet VEV, namely  $v_T\lsim5$ GeV.  As emphasized in \cite{PBAS,PBAS2,PBAS3} a non-zero triplet VEV can have drastic impact on the Higgs sector and possibly on other sectors.  Throughout this paper $v_T$ is fixed at $3\sqrt{2}$ GeV. 

Superpartners of  the $SU(2)$ doublet and triplet Higgs bosons and of $W$ and $B$ bosons constitute the neutralinos and chargino sectors. At EWSB their same charge gauge eigenstates mix also with each other. 
The production and decay phenomenology of neutralinos and charginos depend strongly on the mixing angles.  For triplinos (superpartners of triplets), couplings with fermions are proportional to the doublet-triplet mixing angle, since directly triplinos do not couple to the fermions (or sfermions). In the subsections below we discuss the neutralino and chargino sectors separately, as well as the mixings among different gauge states. 

 \subsection{Neutralino sector}
Eq.~\eqref{ntrl} below presents the neutralino mass matrix in the basis ($\tilde{B}^0$, $\tilde{W}^0$, $\tilde{H}_d^0$, $\tilde{H}_u^0$, $\tilde{T}^0$). It is seen that the triplet VEV $v_T$ does not generate any independent 
mixing, but rather takes part in the bi-linear Higgsino mixing between $\tilde{H}_u^0$ and $\tilde{H}_d^0$. Triplino mixes with $\tilde{H}_u^0$ and $\tilde{H}_d^0$ via the coupling $\lambda$ and does not mix at all 
with $\tilde{B}^0$, $\tilde{W}^0$. This is due to the fact that $T_0$ component does not couple to the $T_3=0, Y=0$ states, which makes wino/bino like NLSP/LSP with a triplino LSP/NLSP scenarios particularly interesting.  
\begin{eqnarray}\label{ntrl}
\mathcal{M}_{\tilde{\chi}^0}=
\left(
\begin{array}{ccccc}
M_1\;\; & 0\;\; & -\frac{1}{2} g_Y v_d\;\; & \frac{1}{2} g_Y v_u\;\; & 0 \\
0\;\; & M_2\;\; & \frac{1}{2} g_L v_d\;\; & -\frac{1}{2} g_L v_u\;\; & 0 \\
-\frac{1}{2} g_Y v_d\;\; & \frac{1}{2} g_L v_d\;\; & 0\;\; & -\mu _D+\frac{1}{2}\lambda  v_T \;\;&
\frac{1}{2} \lambda  v_u\\
\frac{1}{2} g_Y v_u\;\; & -\frac{1}{2} g_L v_u\;\; & -\mu _D+\frac{1}{2} \lambda  v_T\;\; & 0\;\; &
\frac{1}{2} \lambda  v_d \\
0\;\; & 0\;\; & \frac{1}{2} \lambda  v_u\;\; & \frac{1}{2}\lambda  v_d\;\; & 2 \mu_T \\
\end{array}
\right). 
\end{eqnarray}
%
The neutral parts of the $Y=0$ triplet superfield, the triplet scalar or triplino, do not couple to the $Z$ boson. Thus $\ntrlj \to Z \ntrli$ is not allowed for a pure triplet-like neutralino. Such inertness of the triplino decays is important especially for triplino-like NLSP or LSP.   Triplet-like neutralino (NLSP) decays with a small mass gap with LSP ($\sim 100-200$ GeV) can be essential in determining the nature of the LSP or the possible mixing among the neutralinos.  In particular the $\ntrlj \to Z/h/A \ntrli$ and $\ntrlj \to W^\pm/H^\pm \chi^\mp_i$ decays are crucial in determining the characteristics of the neutralinos i.e., triplet-doublet-bino-wino types of neutralinos.

\subsection{Chargino Sector}
For the chargino sector, the mass matrix appears in the Lagrangian with the three column vectors $\psi^+=$($\tilde{W}^+$,$\tilde{H}_u^+$,$\tilde{T}_2^+$) and $\psi^-=$($\tilde{W}^-$,$\tilde{H}_d^-$,$\tilde{T}_1^-$) as 
\begin{eqnarray}
\mathcal{L}\supset - (\psi^-)^{T} \mathcal{M_{\tilde{\chi}^\pm}} \psi^+ + h.c\ ,
\end{eqnarray}
where 
\begin{eqnarray}
M_{\tilde{\chi}^{\pm}}=\left(
\begin{array}{ccc}
 M_2\;\; & \frac{1}{\sqrt{2}}g_L v_u\;\; & -g_L v_T \\
 \frac{1}{\sqrt{2}}g_L v_d\;\; & \frac{1}{2} \lambda  v_T+\mu _D\;\; & \frac{1}{\sqrt{2}}\lambda  v_u
   \\
 g_L v_T\;\; & \frac{-1}{\sqrt{2}}\lambda  v_d\;\; & 2 \mu _{T} \\
\end{array}
\right).  
\end{eqnarray}
Unlike the triplet-like neutralino, the triplet-like chargino mixes with both Higgsino- and wino-like charginos via $g_L v_T$ and $\lambda  v_{u/d}$. Interestingly for an inert triplet, {\it i.e.} $v_T=0$, the wino-triplino mixing 
vanishes, however, Higgsino-triplino mixings remain, and they go to zero only in the decoupling limit $\lambda =0$. Unlike the neutral triplinos, the charged triplinos, being charged under $SU(2)$, couple to 
$W^\pm, \tilde{W}^\pm$ and $W^0, \tilde{W}^0$.  However, similar to the neutralino sector this type of charginos do not couple to fermions or sfermions. This feature affects the bounds from the rare decays 
like $B \to X_s \gamma$ \cite{PBAS2}. In the next section we scan the parameter space to look for such states and the corresponding phenomenology. 

\section{Parameter space scan}\label{[param]}
For the phenomenological analysis at the LHC with center of mass energy of 14 TeV, we look for the suitable parameter space. For this purpose we scan the parameter space in the regions defined by
\begin{eqnarray}\label{pscan}
&&1\leq {\tan\beta }\leq 50,\  \left|\lambda\right|\leq1.2,\,0\leq \left|\mu _D,\mu _T\right|\leq 2 \,\text{TeV},\ 100 \,\text{GeV}\leq \left|M_1,M_2\right|\leq 2  \,\text{TeV},\nonumber\\ 
&& 0\leq \left| A_t,A_b,A_{\lambda},B_D,B_T\right|\leq 2 \,\text{TeV},\ 500 \,\text{GeV}\leq m_Q,m_{\tilde{t}},m_{\tilde{b}}\leq 2\,\text{TeV} ,\end{eqnarray}
where $m_Q,m_{\tilde{t}},m_{\tilde{b}}$ are the left- and right-handed squark soft masses and $M_i$ (i=1,2) are the soft gaugino masses. The parameter scan is performed for various scenarios differing in neutralino/chargino natures. Apart from satisfying the individual requirements for corresponding scenarios, each collected data point respects the following constraints: 
\begin{eqnarray}\label{cons}
&&124.6\leq\ m_{h_1^0}\leq 125.6\, {\rm GeV}\ ;\ m_{A_{1,2}},\ m_{\tilde{\chi}^0_{1,2,3,4,5}}\geq  65\,{\rm GeV}\ ;\nonumber\\
 &&\,m_{\tilde{\chi}^\pm_{1,2,3}}\geq 104\,{\rm GeV} \ ; \ m_{\tilde{t}_{1,2}},m_{\tilde{b}_{1,2}}>  600\,{\rm GeV}\ .
\end{eqnarray}
In Eq.~\eqref{cons} we consider the current Higgs data \cite{Higgsd1, Higgsd2} for $\sim 125$ GeV Higgs. We avoid the invisible decay width of the $\sim 125$ GeV Higgs boson by demanding 
$\ m_{A_{1,2}},\ m_{\tilde{\chi}^0_{1,2,3,4,5}}\geq 65$ GeV. The most recent bounds on the third generation squarks are considered \cite{thirdgen} along with the bounds from the electroweak charginos and 
neutralinos \cite{winoNLSP, winoNLSPATLAS}. In the following subsection we construct three different scenarios by having different nature of the NLSP and LSP. Later in this article we explore the possibilities of having long-lived triplinos and displaced vertices with such parameters at the LHC.

\subsection{Sc 1: Triplino LSP}
\begin{figure}[tbh]
	\begin{center}
		\includegraphics[width=0.6\linewidth]{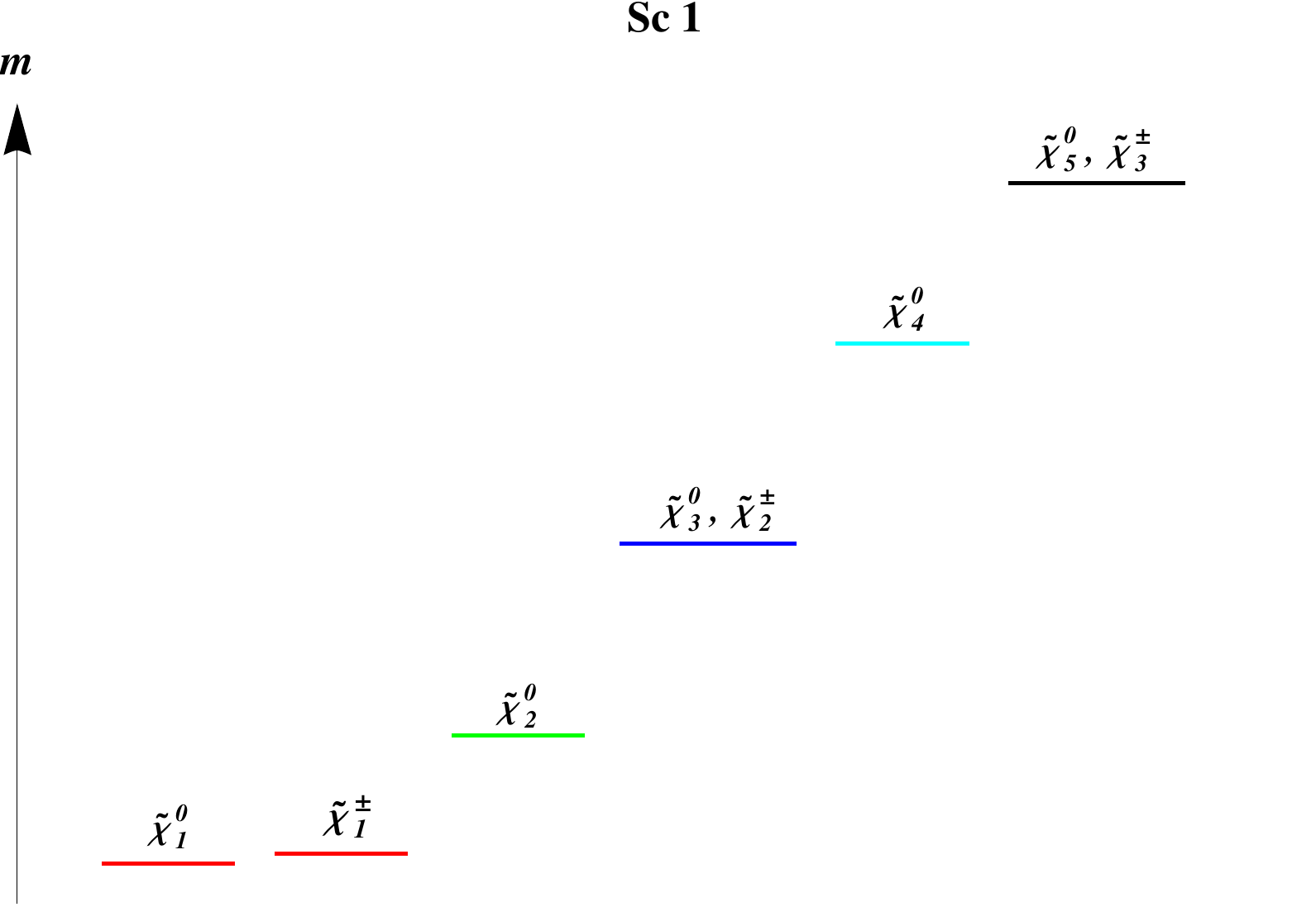}
		\caption{Mass hierarchy among neutralinos and charginos for Sc 1, where we have triplino-like LSP and NLSP and bino-like NNLSP. }\label{massd_ScI}
	\end{center}
\end{figure}
\begin{figure}[tbh]
\mbox{\subfigure[]{\includegraphics[width=70mm,scale=1]{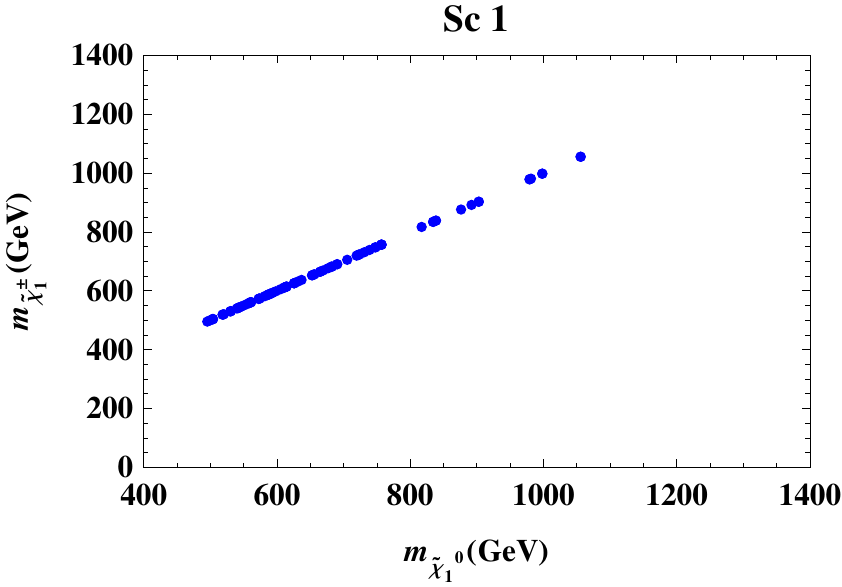}}
\subfigure[]{\includegraphics[width=0.5\linewidth]{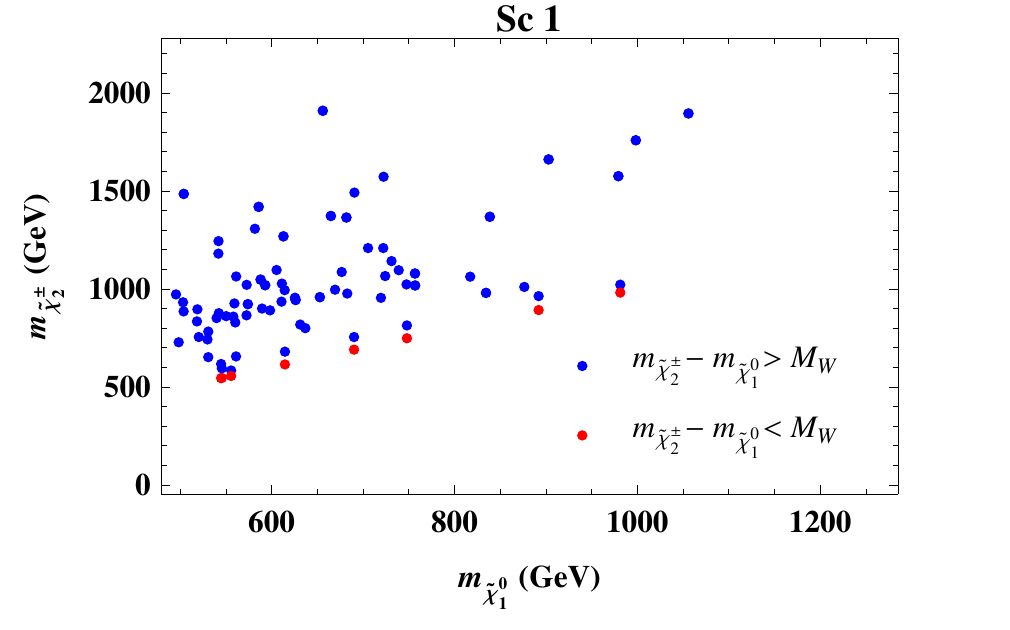}}}
\caption{In Sc 1: (a) LSP mass versus the lightest chargino $\tilde{\chi}_1^{\pm}$, and (b)  LSP mass versus the second lightest chargino $\tilde{\chi}_2^{\pm}$.}\label{nu1ch12_sc1}
\end{figure}

 For Sc 1 we choose the data points that respect the common constraints in Eq. \eqref{cons}  and contain more than $90\%$ triplino-like LSP in the spectrum. For the obtained data points of Sc 1 we display the mass 
 hierarchy among  charginos and neutralinos in Fig. \ref{massd_ScI}. Here the red colour corresponds to triplet-like chargino and neutralinos, the mostly bino-like NNLSP is in green colour. The heavier charginos and 
 neutralinos arrange themselves as either degenerate wino or Higgsino states, and they are shown in blue and black colour. One of the neutralinos is not degenerate with any other state, shown with cyan colour.  We observe that the lightest chargino $\tilde{\chi}_1^{\pm}$ is the NLSP and it is nearly mass-degenerate with LSP as shown in Fig.\ref{nu1ch12_sc1} (a).  For such a mass degeneracy,  one-loop order neutralino and chargino masses must be taken into account to see whether the degeneracy is preserved after including the quantum corrections \cite{Chen:1996ap, strumia, ejc1}. If such a small mass difference persists it can provide interesting signatures, since the suppressed phase space allows the chargino NLSP to travel some distance before decaying to daughter particles. For the decay width of chargino NLSP  $\leq \mathcal{O}(10^{-13})$ GeV,  it can decay to pions with very small momentum or displaced charged lepton via three body decays depending on the mass difference.  When the mass difference between chargino NLSP and LSP is $\mathcal{O}(150)$ MeV, $\tilde{\chi}_1^{\pm} \to \pi^+ \tilde{\chi}_1^{0}$ mode is open and it becomes the dominant decay mode.  The emitted pion typically has very low momentum and it is not constructed in the detector. Thus the  chargino NLSP just leaves a disappearing track in the detector \cite{disch}.  The long-lived chargino NLSP appearing as a disappearing track is well expected in our scenario, since the decay width of triplino-like chargino NLSP to triplino-like LSP plus  fermions is strongly suppressed due to the lack of triplino-fermion coupling. Fig.\ref{nu1ch12_sc1} (b) shows the mass hierarchy between the LSP and the second lightest chargino. The red points correspond to their mass gap less than $m_W$, making $\chmp2$ eligible only for three-body decays and blue points correspond to a mass gap greater than $m_W$.

\begin{figure}[tbh]
\mbox{\subfigure[]{\includegraphics[width=0.5\linewidth]{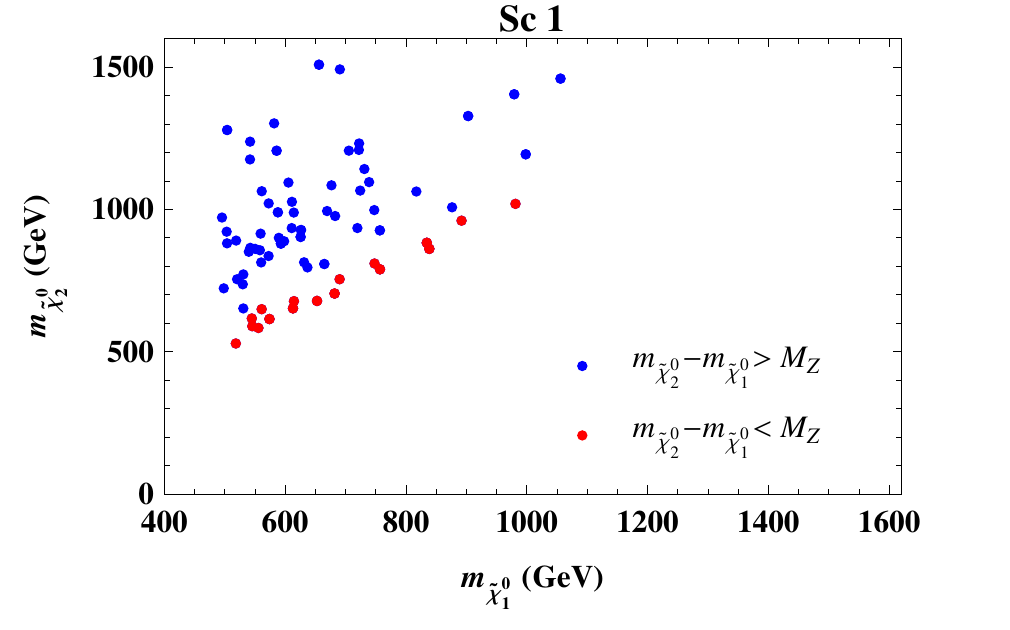}}
\subfigure[]{\includegraphics[width=0.5\linewidth]{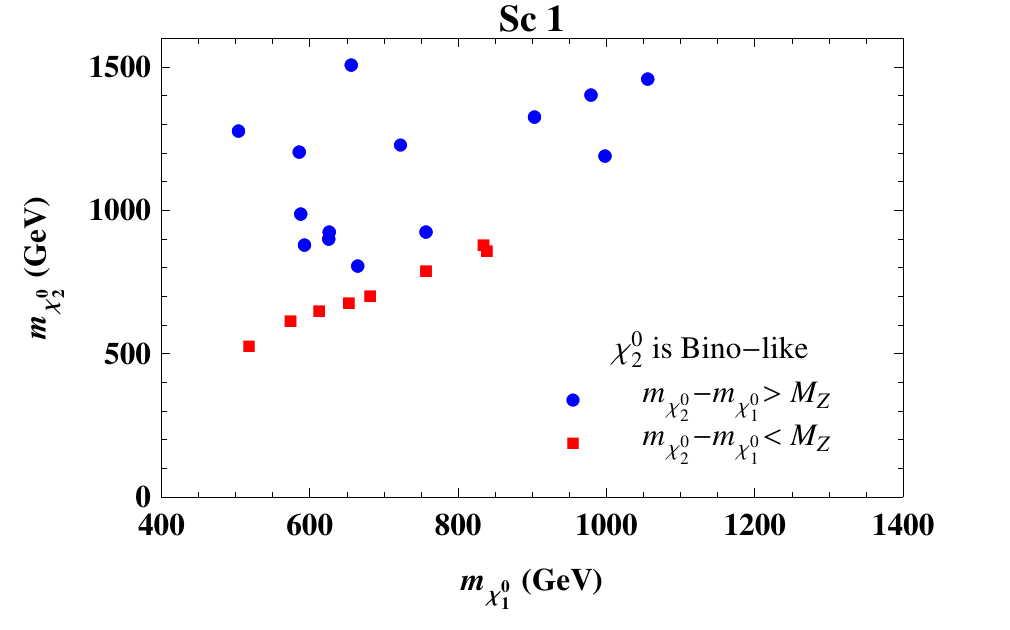}}}
\caption{In Sc 1: (a) LSP versus the second lightest neutralino  $\tilde{\chi}_2^0$ where LSP is mostly triplino type, and (b)  $\tilde{\chi}_2^0$ is bino-like.}\label{nu1nu2_sc1}
\end{figure}

We also investigate gauge modes  through the second lightest neutralino  decay and  the data points with $m_{\tilde{\chi}_2^{0}}-m_{\tilde{\chi}_1^{0}} < M_Z$, which are marked with red in Fig.~\ref{nu1nu2_sc1}. As pointed out earlier,  the LSP production via gauge mode $\tilde{\chi}_2^{0}\rightarrow \tilde{\chi}_1^{0}  Z$  is kinematically impossible for red data points and 3-body decays must be investigated.  In Fig. \ref{nu1nu2_sc1} (b) we investigate the scenario in more detail and we find out that most data points are with bino-like second lightest neutralino. This particular scenario can be interesting in terms of displaced vertices since the absence of bino-triplino coupling results in suppressed $\tilde{\chi}_2^0 \tilde{\chi}_1^0$ coupling so that the second lightest neutralino can live long enough and lead to displaced leptons. However, for the benchmark points taken from this scenario we did not find a noticeable displaced decays for $\ntrl2$. 

 \subsection{Sc 2: Triplino NLSP}
\begin{figure}[tbh]
	\begin{center}
\includegraphics[width=0.6\linewidth]{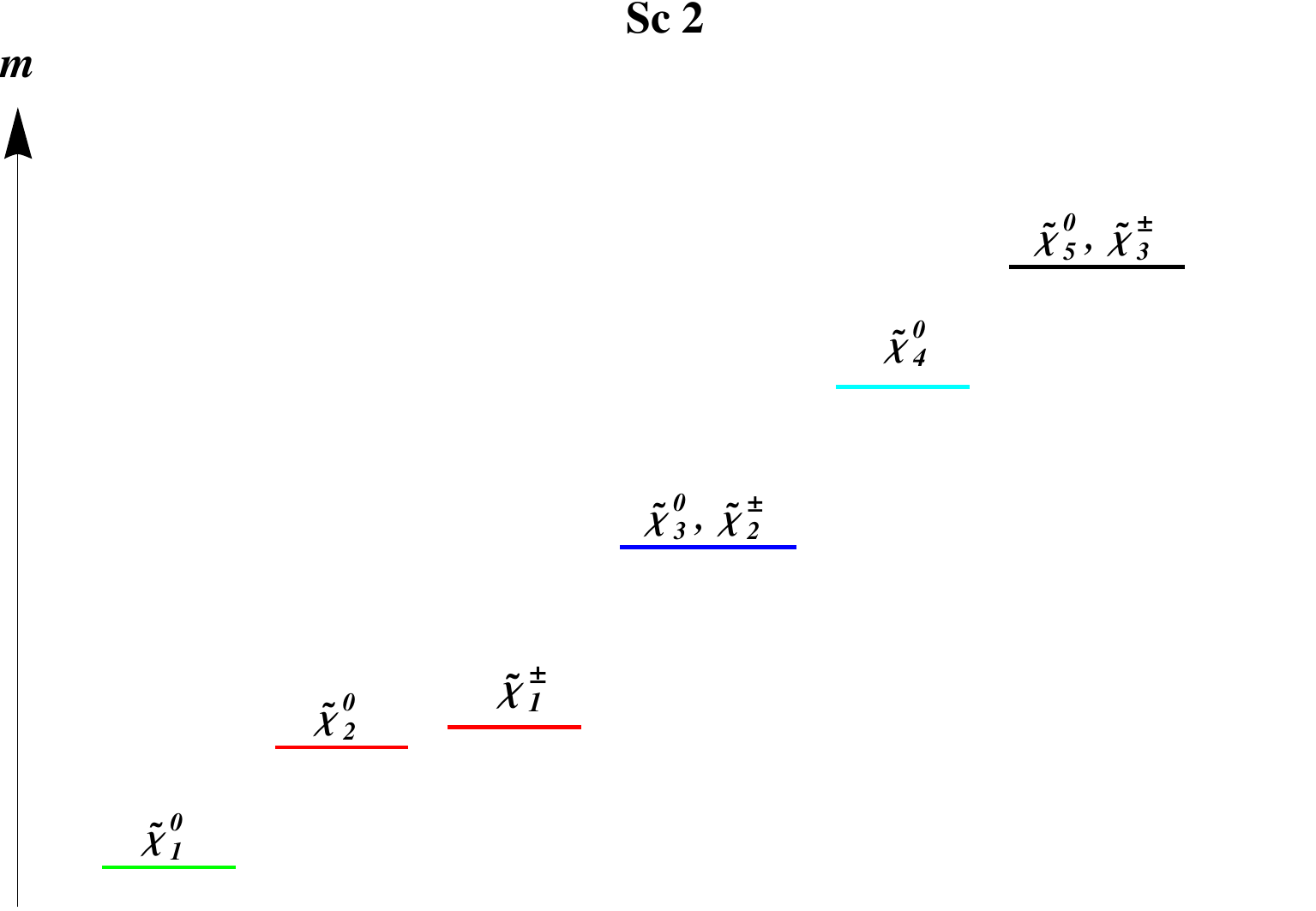}
\caption{Mass hierarchy among neutralinos and charginos for Sc 2 with a triplino-like second lightest neutralino.}\label{massd_ScII}
	\end{center}
\end{figure}

In Sc 2 we focus on the phenomenology of the triplino like second lightest neutralino $\tilde{\chi}_{2}^{0}$ chosen as NLSP. We have performed a scan respecting the constraints given in Eq. \eqref{cons} and ask for the 
points where triplino component of NLSP is more than $90\%$. This scenario is quite interesting, since for the triplino-like NLSP the current LSP mass bound from the lepton mode 
$\tilde{\chi}_{2}^{0}\rightarrow  \ell \tilde{ \ell} \rightarrow \ell \ell \tilde{\chi}_{1}^{0}$  is less tight \cite{winoNLSP, winoNLSPATLAS} because NLSP does not couple to leptons due to its triplino nature. The mass hierarchy 
among neutralinos and charginos for the data points is given in Fig. \ref{massd_ScII}. Here the red colour corresponds to the triplet-like chargino and neutralinos, the bino-like LSP is denoted by green colour. The heavier 
charginos and neutralinos arrange themselves also in Sc 2 as either degenerate wino or Higgsino states, and are denoted by blue and black colour. Again one of the neutralinos is not degenerate with other particles, and is shown with cyan colour.
\begin{figure}[b]
	\mbox{\subfigure[]{\includegraphics[width=0.5\linewidth]{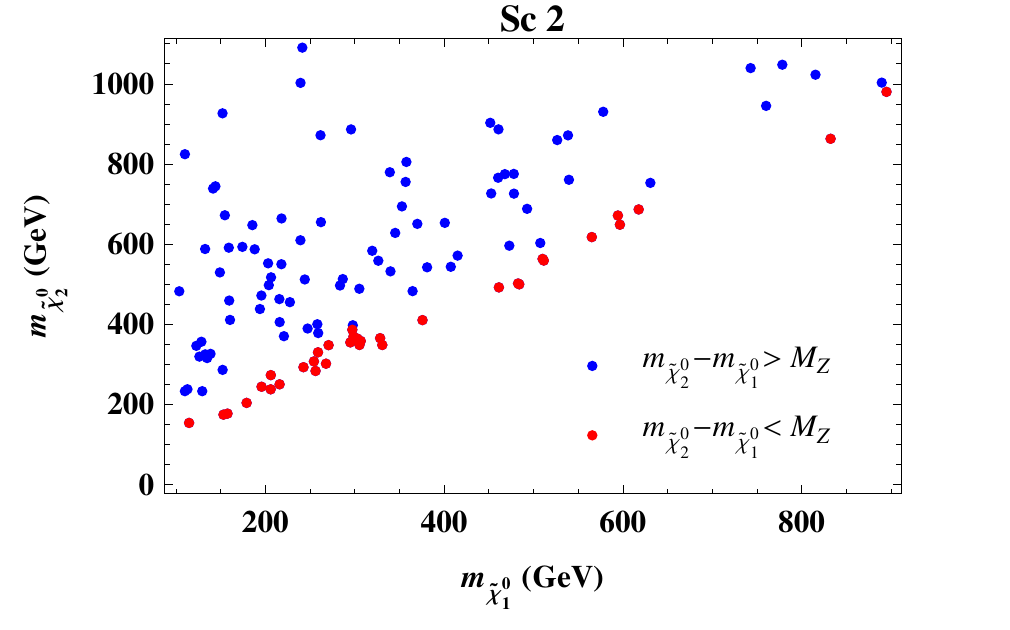}}
		\subfigure[]{\includegraphics[width=0.5\linewidth]{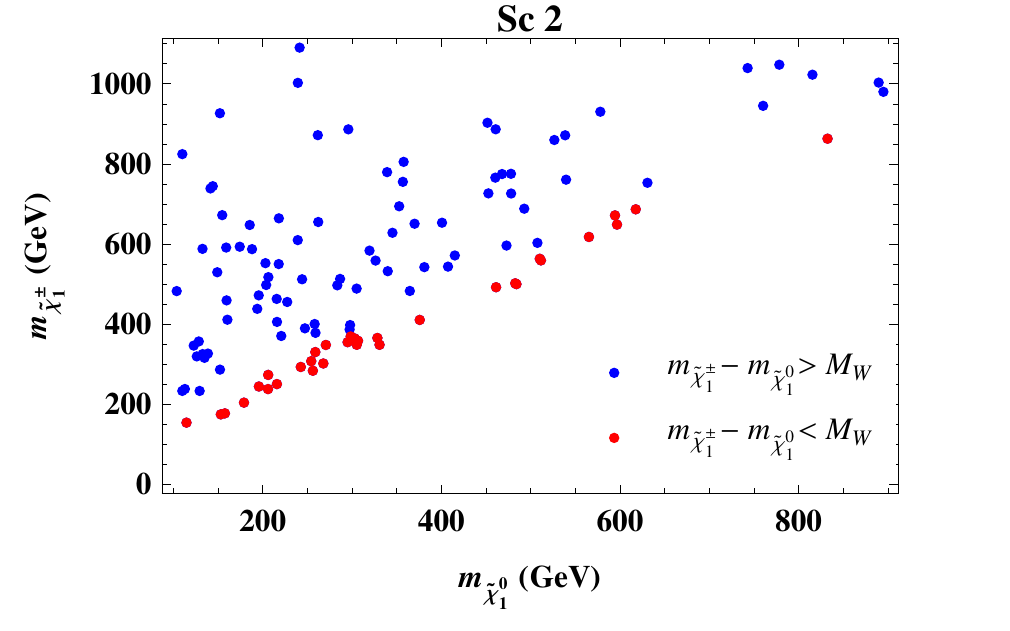}}}
	\caption{In Sc 2: bino-like LSP versus (a) triplino-like NLSP, and  (b) the lightest chargino, where mass difference greater/less than $M_Z$ ($M_W$) is colour-coded.}\label{nu1nu2_sc2}
\end{figure}
The most striking feature of this scenario is that  requiring a triplet like neutralino NLSP leads to  bino-like LSP in all data sets.  
This is because the charged and neutral states with same nature are almost mass degenerate, and thus requirement of a triplino-like neutralino NLSP leads to a  chargino with mass always slightly greater than the corresponding neutralino in the same gauge representation.
Demanding either wino- or triplino-like neutralino NLSP leaves no choice but bino to be LSP, since bino does not have any charged partner.  
The triplino-like neutralino NLSP leads to the lightest chargino with mass very close to NLSP making it next to NLSP (NNLSP).

In Fig. \ref{nu1nu2_sc2} (a) we  display the mass hierarchy between LSP and NLSP where mass difference less than $M_Z$ is red colour-coded. For the $m_{\tilde{\chi}_2^{0}}-m_{\tilde{\chi}_1^{0}} > M_Z$ case we expect $\tilde{\chi}_2^{0}\tilde{\chi}_1^{0}Z$ coupling to be strongly suppressed due to the absence of triplino-$Z$ -coupling. 
In Fig. \ref{nu1nu2_sc2} (b) we display the mass difference between the lightest chargino and LSP. It can be less than $M_W$ for some data points, yet for many data points $\tilde{\chi}_1^{\pm}\rightarrow \tilde{\chi}_1^{0}  W^{\pm}$ decay is kinematically possible.

\subsection{Sc 3: Higgsino LSP}
\begin{figure}[tbh]
	\begin{center}
		\includegraphics[width=0.6\linewidth]{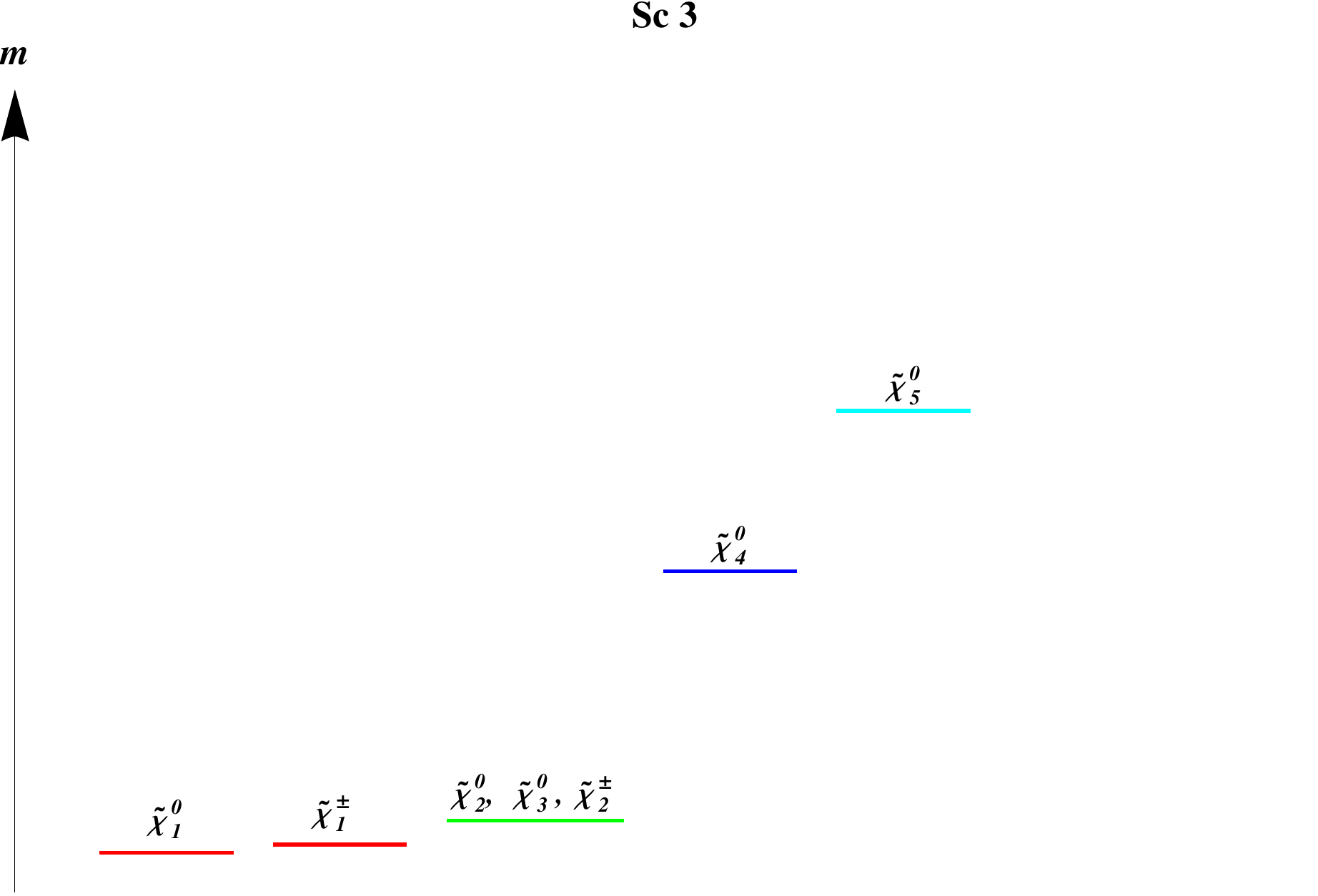}
		\caption{Mass hierarchy among neutralinos and charginos for Sc 3. The heaviest chargino ($\tilde{\chi}_3^{\pm}$) is mass degenerate either with $\tilde{\chi}_4^{0}$ or $\tilde{\chi}_5^{0}$ depending on the fields dominating its structure. }\label{massd_ScIII}
	\end{center}
\end{figure}

In the search of long lived neutralinos and charginos  we also dwell on the possibility of having Higgsino-like LSP and triplino-like NLSP, whose interaction vertex is proportional to trilinear coupling $\lambda$. For small values of  
$\lambda$, strongly triplino-like NLSP can be quite long lived before decaying to Higgsino dominated LSP  and SM particles. To investigate this possibility we demand that the LSP is at least 50 $\%$ Higgsino-like and 
the NLSP is at least 50 $\%$ triplino-like during the parameter scan of scenario Sc 3.  In Fig.~\ref{massd_ScIII} we observe that the NLSP turns out to be the lightest chargino which is almost mass degenerate 
and has the same Higgsino-like nature with the LSP. In this respect  this scenario is similar with  Sc 1 and one needs to calculate the one-loop masses for neutralinos and charginos to see  if quantum corrections can change the mass hierarchy \cite{strumia, ejc1}.

\begin{figure}[tbh]
	\mbox{\subfigure[]{\includegraphics[width=0.51\linewidth]{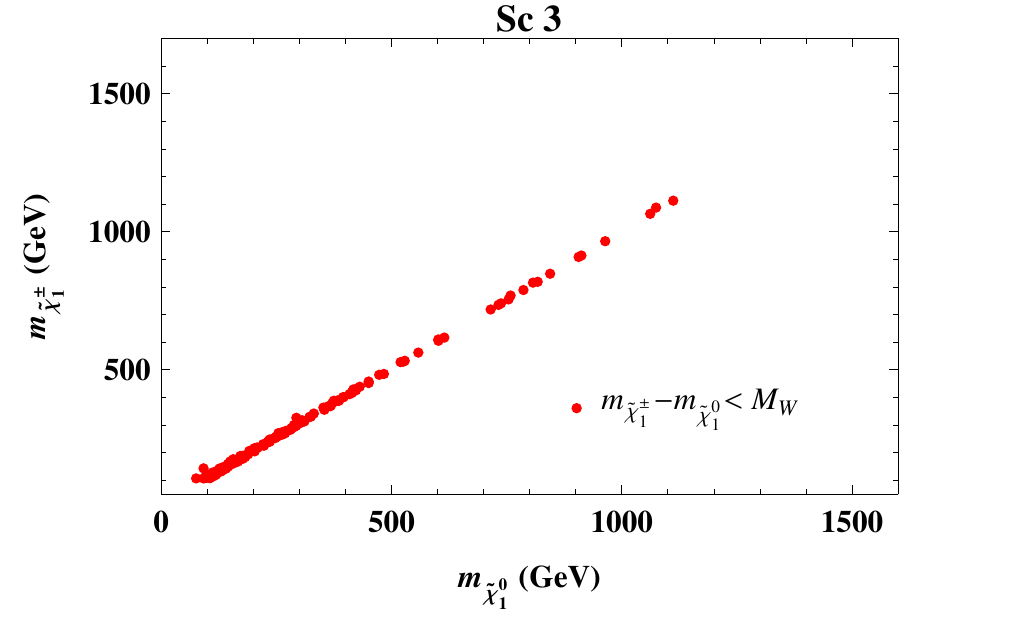}}
		\subfigure[]{\includegraphics[width=0.51\linewidth]{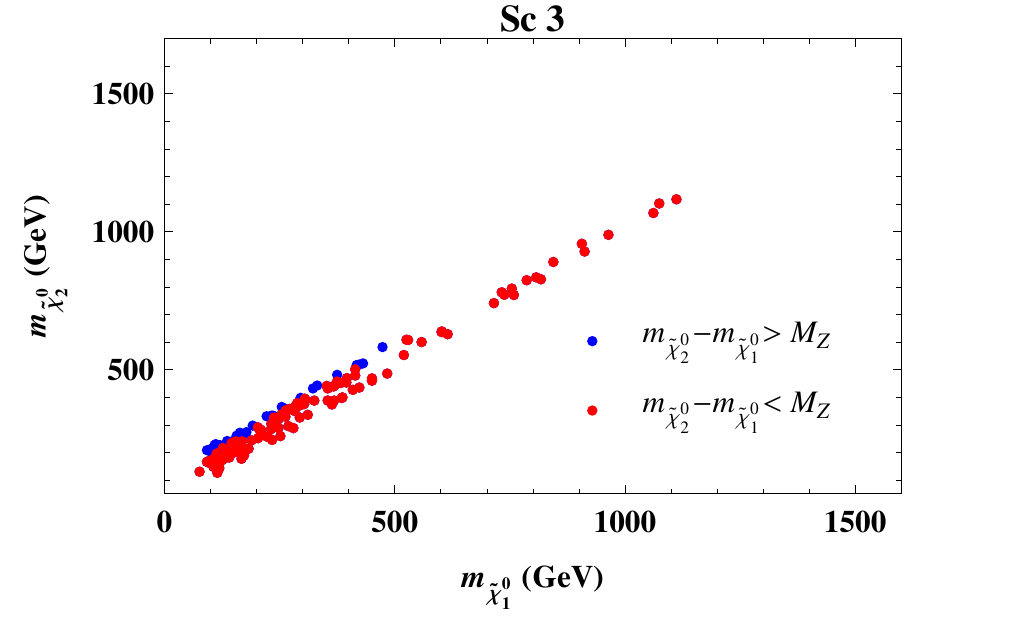}}}
	\caption{In Sc 3: LSP versus (a) chargino NLSP and  (b) neutralino NNLSP where  $m_{\tilde{\chi}_2^{0}}-m_{\tilde{\chi}_1^{0}} > M_Z$ data points are  coloured in blue.}\label{nu1nu2_sc3}
\end{figure}

We also notice that  the second lightest neutralino is NNSLP and it is nearly mass degenerate with the second lightest chargino $\tilde{\chi}_2^{\pm}$ and the third lightest neutralino $\tilde{\chi}_3^{0}$. The most distinctive difference 
compared to the other two scenarios is that Sc 3 contains four neutralinos and charginos having mass values close to the LSP mass. Fig.~\ref{nu1nu2_sc3} (a) shows the mass hierarchy between the chargino NLSP and 
neutralino LSP and they are almost degenerate.  In Fig.~\ref{nu1nu2_sc3} (b) we show the mass hierarchy between neutralino NNLSP and LSP and the data points with the mass difference greater than $M_Z$  are shown 
in blue and such points are less in number.  For the data points for which $\tilde{\chi}_2^{0}\rightarrow \tilde{\chi}_1^{0} Z (h)$ is not kinematically possible, the 3-body decay channels must be investigated for LSP 
production. The following section is dedicated to the search for the displaced decays at the LHC via a PYTHIA based simulation \cite{pythia}.

\section{LHC phenomenology}\label{pheno}
In this section we look for the displaced tracks for the scenarios discussed above by selecting few benchmark points from each scenario. Before going into detailed analysis we first introduce our set up for the simulation at the LHC with 14 TeV of center of mass energy. 
After the set up we discuss the phenomenology of the different scenarios separately. 

To start  we implemented the model in SARAH \cite{sarah1, sarah2} then generated the model files for CalcHEP \cite{calchep}, which has been used to generate the 'lhe events' containing the decay branching ratios and the corresponding mass spectra. The generated events have then been simulated with {\tt PYTHIA} \cite{pythia} for hadronization and showering via 'lhe' interfacing \cite{lhe}.The simulation at hadronic level has been performed using the {\tt Fastjet-3.0.3} \cite{fastjet} with the {\tt CAMBRIDGE AACHEN} algorithm. We have selected a jet size $R=0.5$ for the jet formation, with the following criteria:

\begin{itemize}
	\item the calorimeter coverage is $\rm |\eta| < 4.5$
	
	\item the minimum transverse momentum of the jet $ p_{T,min}^{jet} = 10$ GeV and jets are ordered in $p_{T}$
	\item leptons ($\rm \ell=e,~\mu$) are selected with
	$p_T \ge 5$ GeV and $\rm |\eta| \le 2.5$
	\item no jet should be accompanied by a hard lepton in the event
	\item $\Delta R_{lj}\geq 0.4$ and $\Delta R_{ll}\geq 0.2$
	\item Since an efficient identification of the leptons is crucial for our study, we additionally require  
	a hadronic activity within a cone of $\Delta R = 0.3$ between two isolated leptons to be $\leq 0.15\, p^{\ell}_T$ GeV, with 
	$p^{\ell}_T$ the transverse momentum of the lepton, in the specified cone.
	
\end{itemize}

\subsection{Sc 1: Triplino LSP}
\begin{table}
	\begin{center}
		\renewcommand{\arraystretch}{1.4}
		\begin{tabular}{||c|c|c|c|c|c||}
			\hline\hline
			Benchmark&LSP mass &NLSP mass &NNLSP mass & $\tau_{NLSP}$ &$c\, \tau_{NLSP}$\\
			Points & (GeV)&(GeV)& (GeV)&(ns)& (cm)\\ \hline\hline
			BP1 & 542.30 & 542.50 & 864.70 & 0.79   & 23.61\\
			\hline
			BP2 & 561.12 & 561.54 & 651.82 & 0.022 & 0.646 \\
			\hline
			BP3    & 530.75 & 530.94 & 771.40 &  1.21 & 36.15 \\
			\hline
			BP4 & 498.38 & 498.53 & 722.40 & 3.27 & 97.76\\
			\hline
		\end{tabular}
		\caption{Benchmark points from Sc1 for collider study consistent with the $\sim 125$ GeV Higgs mass where the lifetime of NLSP is given as $\tau_{NLSP}$ and the proper decay length of NLSP is given as $c\,\tau_{NLSP}$.   }\label{bpsc1}
	\end{center}
\end{table}
For the LHC simulation we first consider Sc 1, where we have a triplino-like LSP and a triplino-like chargino NLSP. For the collider study we select four benchmark points from this scenario as given in Table~\ref{bpsc1}, where the mass spectra for NNLSP, NLSP and LSP are listed and we can see that NLSP and LSP are nearly degenerate. For all four points NLSP is triplino-like chargino and it can be seen that the decay length for the chargino NLSP is $\mathcal{O}(1-100)$ cm. However, loop corrections can alter the mass hierarchy in which case we can have an electromagnetically charged LSP, {\it i.e.} a dark matter candidate, which is not physical \cite{strumia, ejc1,ejc2}. For this purpose we have checked the mass hierarchy via SPheno \cite{spheno} considering contributions from all particles at one-loop level and the hierarchy remains the same  for all benchmark points under consideration.  From Table~\ref{bpsc1} we see that the tree-level mass difference between the NLSP and LSP  is around $\mathcal{O}(200)$ MeV, sufficient to have $\chpm1 \to \pi^\pm \ntrl1$ decay. 

Table~\ref{no1} shows the pair and associated production cross-sections for the lighter charginos and neutralinos at the LHC with center of mass energy of 14 TeV. The renormalization and factorization scales are chosen to be $\hat{s}$, and CTEQ6L \cite{6teq6l} is chosen as PDF.  It is evident due to triplino nature of the chargino NLSP and neutralino LSP that the corresponding $\chpm1\ntrl2$ production cross-sections are suppressed by $\mathcal{O}(10^2)$ compared to the wino-like NLSP scenarios \cite{crosswg}. Similarly the other production modes are also suppressed due to the triplino nature of the lighter neutralino and chargino states. The decay products are also soft such that they would be missed even at trigger level. This is why such processes could not be probed with the data from LHC at 8 TeV, and thus no mass limits can be drawn from 8 TeV data  \cite{winoNLSP, winoNLSPATLAS}. 
\begin{table}
	\begin{center}
		\renewcommand{\arraystretch}{1.4}
		\begin{tabular}{||c|c|c|c|c|c|c|c||}
			\hline\hline
			Benchmark&$\tilde{\chi}_1^{\pm}\tilde{\chi}_1^{\mp}$ &$\tilde{\chi}_1^{\pm}\tilde{\chi}_1^{0}$& $\tilde{\chi}_1^{\pm}\tilde{\chi}_2^{0}$&$\tilde{\chi}_1^{\pm}\tilde{\chi}_2^{\mp}$&$\tilde{\chi}_2^{0}\tilde{\chi}_2^{0}$&$\tilde{\chi}_2^{0}\tilde{\chi}_1^{0}$ &$\tilde{\chi}_2^{0}\tilde{\chi}_2^{\pm}$\\
			Points & (fb) &(fb)&(fb)&(fb)&(fb)&(fb)&(fb) \\ \hline
			BP1 &  14.33 & 30.17& 4.71$\times 10^{-2}$ & 0.204  & 1.6$\times 10^{-2}$ & 3.8$\times 10^{-2}$ & 0.82\\
			\hline
			BP2 & 11.50 & 20.03 & 0.12  & 0.352 & 1.4$\times 10^{-6}$  & 6.3$\times 10^{-2}$&  0.89\\
			\hline
			BP3 & 15.79  & 31.97 & 0.56 & 6.0$\times 10^{-2}$  & 8.5$\times 10^{-5}$ & 4.7$\times 10^{-4}$ & 1.17\\
			\hline
			BP4 & 20.84 &  43.71  & 8.37$\times 10^{-2}$  & 7.2$\times 10^{-2}$  & 5.5$\times 10^{-5}$ & 3.74$\times 10^{-4}$& 2.21\\
			\hline
	\hline
\end{tabular}
\caption{ Pair and associated production cross sections for $\tilde{\chi}_{1,2}^{\pm}$ and $\tilde{\chi}_{1,2}^{0}$ at 14 TeV for the benchmark points for Sc I. }\label{no1}
\end{center}
\end{table}

\begin{figure}[tbh]
	\begin{center}
		\subfigure[]{\includegraphics[width=0.3\linewidth, angle=-90]{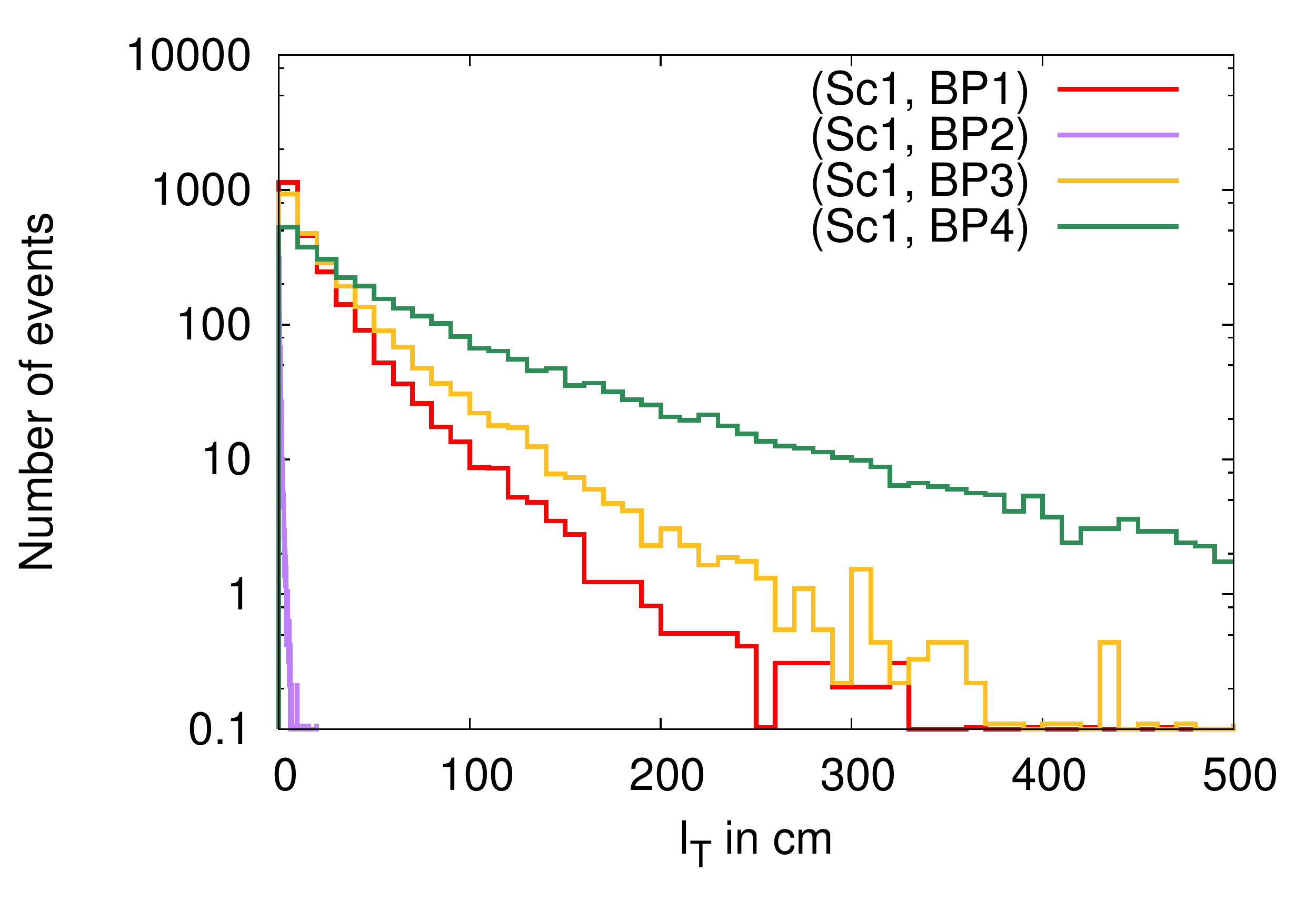}}
		\subfigure[]{\includegraphics[width=0.3\linewidth, angle=-90]{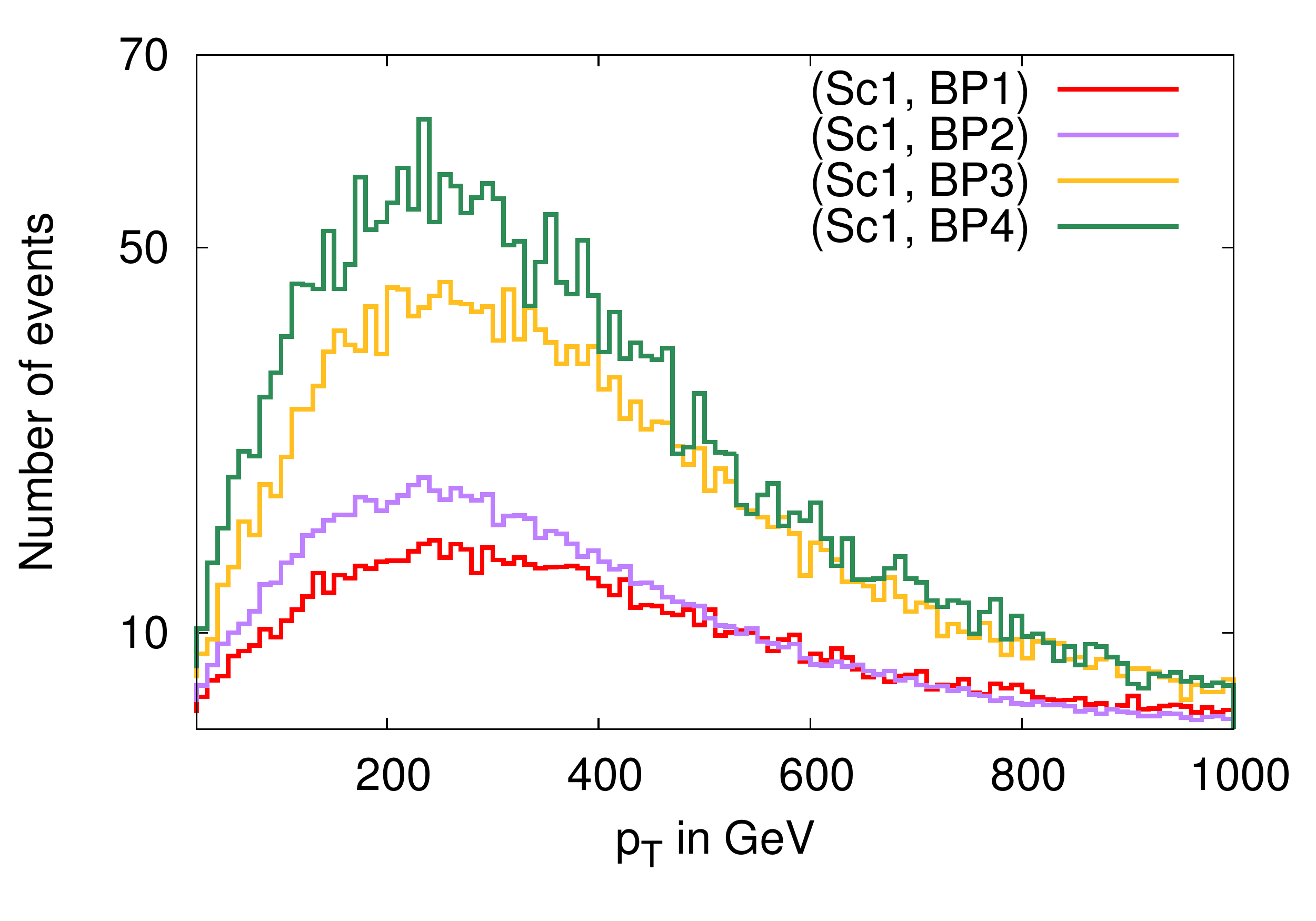}}
		\subfigure[]{\includegraphics[width=0.4\linewidth, angle=-90]{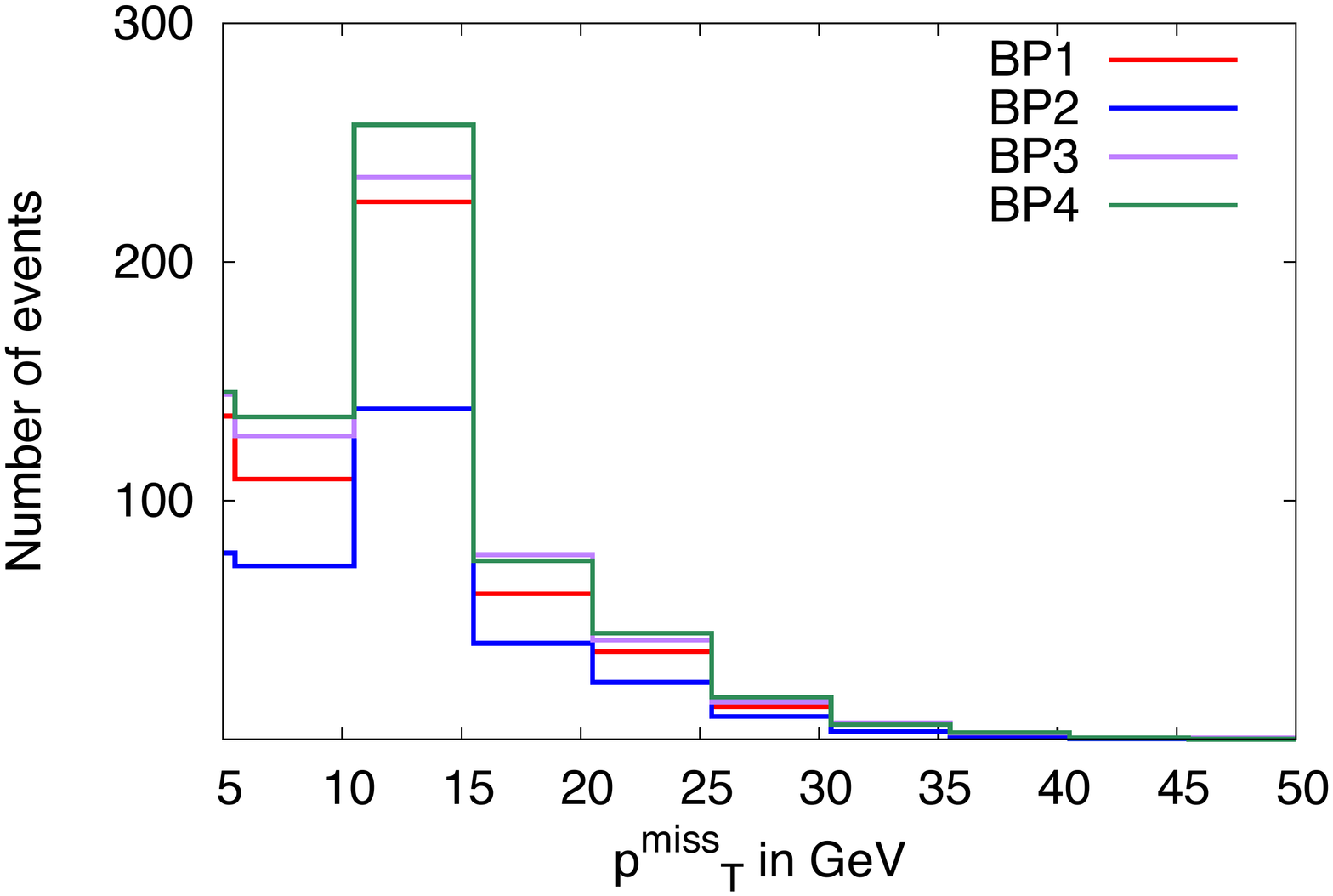}}
		\caption{ The distributions for  the (a) transverse displaced decay length, (b) $p_T$ of the chargino NLSP  and   (c) missing $p_T$  distribution for the benchmark points in Sc1 at the LHC with 14 TeV.}\label{dplsc1}
	\end{center}
\end{figure}

\begin{table}
	\begin{center}
		\renewcommand{\arraystretch}{1.1}
		\begin{tabular}{||c|c|c|c|c|c||}
			\hline\hline
			Benchmark&Process &BP1&BP2& BP3&BP4\\
			Points & &&&&  \\ \hline\hline
			\multirow{2}{*}{0.1-1cm} &$\tilde{\chi}_1^{\pm}\tilde{\chi}_1^{\mp}$& 119.5& 960.3&89.5&46.4\\
			&$\tilde{\chi}_1^{\pm}\tilde{\chi}_1^{0}$&291.1&1715.4&204.7&108.9\\
			\hline
			\multirow{2}{*}{1-10cm} &$\tilde{\chi}_1^{\pm}\tilde{\chi}_1^{\mp}$&599.1&111.8&518.8&329.2\\
			&$\tilde{\chi}_1^{\pm}\tilde{\chi}_1^{0}$&1374.2& 209.6&1141.1&779.2\\
			\hline
			\multirow{2}{*}{10cm -1m}&$\tilde{\chi}_1^{\pm}\tilde{\chi}_1^{\mp}$& 685.5& 0.3&889.8&1194.7\\
			&$\tilde{\chi}_1^{\pm}\tilde{\chi}_1^{0}$&1314.4&0.4&1734.5&2598.1\\
			\hline
			\multirow{2}{*}{1-10m }&$\tilde{\chi}_1^{\pm}\tilde{\chi}_1^{\mp}$& 27.4&0.0&79.9&510.5\\
			&$\tilde{\chi}_1^{\pm}\tilde{\chi}_1^{0}$&34.6&0.0&114.3&881.1\\
			\hline
			\hline
		\end{tabular}
		\caption{ Number of displaced events with disappearing charged track in the ranges of $1-10$ cm, $0.1-1$ m and $1-10$ m for the benchmark points of scenario Sc 1 at the LHC with 14 TeV center of mass energy and at an integrated luminosity of 100 fb$^{-1}$.  }\label{sc1}
	\end{center}
\end{table}
In Fig.~\ref{dplsc1} (a) we plot the transverse decay length of the triplino-like  chargino NLSP for the benchmark points at the LHC. It can be seen that except for BP2 other BPs have $\mathcal{O}(100)$ cm transverse decay lengths. For BP2 it is of the order of cm. 
The transverse momentum distribution for the chargino NLSP is very hard, as shown in Fig.~\ref{dplsc1}(b). However, due to very small mass gap between NLSP and LSP, the decay products are rather soft and would be missed by the LHC detector at the trigger level. This is evident from the missing $p_T$ distributions as shown in Fig.~\ref{dplsc1} (c). The mass spectra for the benchmark points constitute of  heavy LSP  which apparently should give rise to large missing momenta. However due to the compressed spectrum, the NLSPs as well as the LSPs  are produced mostly back to back and there is cancellation among the momenta of LSPs, which results into very low missing momenta. This a signature of nearly degenerate mass spectrum with dark matter candidate and very common in UED \cite{pbad}. However, sufficient boost of the produced neutralino and chargino can give much needed momentum to their decay products. 
A possibly useful strategy is to tag the hard initial state radiation jet, which will lead to relatively large momentum recoil for the desired final state, and the decay products can cross the trigger level cuts \cite{tao}. Such analysis, though mentioned in the literature, has not been performed by LHC experiments so far while looking for electroweak neutralinos and charginos. In the following Sections we will explore these possibilities.

\begin{table}
	\begin{center}
	\renewcommand{\arraystretch}{1.4}
	\begin{tabular}{||c|c|c|c|c||}
		\hline\hline
		Benchmark&\multicolumn{2}{|c|}{$n_{\ell}\geq 1$} &\multicolumn{2}{|c|}{$n_{\ell}\geq 2$}\\
		\cline{2-5}
		Points & $\chpm1\chmp1$&$\chpm1\ntrl1$&$\chpm1\chmp1$&$\chpm1\ntrl1$\\ \hline\hline
		BP1 &2.1	$\begin{cases}
	    0.0\\
		0.3\\    
		0.9\\
		0.9    
		\end{cases}$&3.2$\begin{cases}
		 0.2\\
		0.2\\    
		1.6\\
		1.2    
		\end{cases}$&0.1&0.0\\
		\hline
		BP2 &1.0$\begin{cases}
		0.3\\
		0.6\\    
		0.1\\
		0.0    
		\end{cases}$&55.0$\begin{cases}
	    42.2\\
		12.7\\    
		0.1\\
		0.0    
		\end{cases}$ &0.2&0.0\\
		\hline
		BP3&0.3$\begin{cases}
		0.0\\
		0.0\\    
		0.1\\
		0.2 
		\end{cases}$ &0.0&0.0&0.1\\
		\hline
		BP4&7.7$\begin{cases}
		0.0\\
		0.3\\    
		2.2\\
		5.2 
		\end{cases}$&0.2$\begin{cases}
		0.0\\
		0.0\\    
		0.1\\
		0.1 
		\end{cases}$ &0.0&0.0\\
		\hline
		\hline
	\end{tabular}
	\caption{ Number of events with multiple charged leptons with at least one of them displaced with displacement  0.1 mm  to 10 m at 14 TeV for the benchmark points in scenario Sc 1. Here the leptons are rather soft $p^{\ell}_T \geq 5$ GeV and at least one of them is displaced. }\label{sc12}
\end{center}
\end{table}

\begin{table}
	\begin{center}
		\renewcommand{\arraystretch}{1.4}
		\begin{tabular}{||c|c|c|c|c||}
			\hline\hline
			Benchmark&\multicolumn{2}{|c|}{$n_{j}\geq 1$} &\multicolumn{2}{|c|}{$n_{j}\geq 2$}\\
			\hline
			Points & $\chpm1\chmp1$&$\chpm1\ntrl1$&$\chpm1\chmp1$&$\chpm1\ntrl1$\\ \hline\hline
			BP1 &593.9$\begin{cases}
			50.6\\
			249.4\\    
			 283.9\\
			 10.0 
			\end{cases}$&2455.8$\begin{cases}
			 235.3\\
			1120.4\\    
			1073.1\\
			 27.0   
			\end{cases}$&399.9$\begin{cases}
			34.0\\
			168.9\\    
			190.0\\
			7.0
			\end{cases}$&1819.9$\begin{cases}
			172.1\\
			 826.3\\    
			801.5\\
			20.0  
			\end{cases}$\\
			\hline
			BP2 &878.6$\begin{cases}
			785.7\\
			  92.8\\    
			0.1\\
			0.0    
			\end{cases}$&1589.8$\begin{cases}
			 1413.6\\
			175.9\\    
			0.3\\
			0.0    
			\end{cases}$ &653.8$\begin{cases}
			585.4\\
			68.3\\    
			0.1\\
			0.0    
			\end{cases}$ &1155.6$\begin{cases}
			1030.9\\
			124.5\\    
			0.2\\
			0.0    
			\end{cases}$ \\
			\hline
			BP3& 655.2$\begin{cases}
			37.0\\
			217.2\\    
			 369.1\\
			31.9
			\end{cases}$&2615.5$\begin{cases}
			167.2\\
			928.8\\    
			1424.1\\
			95.4
			\end{cases}$&445.4$\begin{cases}
			26.0\\
			147.8\\    
			249.3\\
			22.3
			\end{cases}$&1938.4$\begin{cases} 125.6\\
			686.9\\
			1054.6\\
			71.3	\end{cases}$\\
			\hline
			BP4&860.9$\begin{cases}
			19.5\\
			138.5\\    
			493.7\\
			209.2
			\end{cases}$
			&3562.0$\begin{cases}
			87.5\\
			634.1\\    
			2118.3\\
			722.1
			\end{cases}$ &577.6$\begin{cases}
			13.3\\
			 89.9\\    
			333.7\\
			 140.7
			\end{cases}$ &2631.9$\begin{cases}
			61.8\\
			468.1\\    
			 1565.7\\
			 536.3
			\end{cases}$ \\
			\hline
			\hline
		\end{tabular}
		\caption{ Number of events with multiple jets with at least one of them displaced with displacement  0.1 mm  to 10 m at 14 TeV for the benchmark points in scenario Sc 1. Here the leptons are rather soft $p^{\ell}_T \geq 5$ GeV and at least one of them is displaced. }\label{sc1j}
	\end{center}
\end{table}

In this scenario the NLSP is a chargino, which has large momentum before decaying into charged leptons and LSP. In Table~\ref{sc1} we show numbers of events which may give rise to disappearing charged tracks before they decay. Such displacement can be as large as 10 m for some benchmark points. It can be seen that the main contribution comes from the lightest chargino pair production ($\chpm1\chmp1$) and the lightest chargino production in association with LSP ($\chpm1\ntrl1$) . Due to degeneracy of the spectrum it is most likely that majority of the charged leptons and jets from the decay remain undetected as they will fall below the initial trigger cuts.Nevertheless, in this Section we try to see such soft charged leptons and jets.

Table~\ref{sc12} presents in Sc 1 the number of charged leptons ($e,\, \mu$) with $p^{\ell}_T \geq 5$ GeV  at the LHC with center of mass energy of 14 TeV at an integrated luminosity of 100 fb$^{-1}$ with one of them 
being produced with a displacement. The displacements can be from 0.1 mm to 10 m as listed in Table~\ref{sc12}. It can be seen that the corresponding leptonic events are only few due to the small branching fraction of 
$\chpm1$ to leptons for the benchmark points. However, for BP2 the  $\chpm1\ntrl1$ contribution have sufficiently many leptonic events due to the relatively large leptonic branching fraction, 
$ \chpm1 \to \ell^\pm \nu \ntrl1 \simeq 37\%$.  Otherwise probing such leptonic final states one has to go for high luminosity LHC $\mathcal{O}(3000)$ fb$^{-1}$.

In Table~\ref{sc1j} we show the displaced jets that come from the decays of chargino-type NLSP for the benchmark points at the LHC with center of mass energy of 14 TeV at an integrated luminosity of 100 fb$^{-1}$. 
Here we demand that at least one of the jets should be displaced and the jet momenta can be rather small $p^j_T \geq 10$ GeV. Due to the isolation criteria (for jet-jet, jet-lepton and lepton-lepton), as given before, the 
contribution is much more from $\chpm1\ntrl1$ than from $\chpm1\chmp1$. Of course the latter has larger cross-section that also adds to the contribution.

\subsection{Sc 2: Triplino NLSP}
Unlike scenario Sc 1, in this case we have  a triplino-like neutralino as NLSP and there is enough mass gap between the NLSP and bino-like LSP, which enhances the possibility of detecting those displaced charged lepton and jets. In Table~\ref{bps2} we list three benchmark points for which the displaced decay length of triplino-like NSLP can be from cm to meter. However, it is interesting to see that the decay branching fractions of such NLSPs are often into $d\bar{d} \ntrl{1}$.

\begin{table}[b]
	\begin{center}
		\renewcommand{\arraystretch}{1.4}
		\begin{tabular}{||c|c|c|c|c|c||}
			\hline\hline
			Benchmark&LSP mass & NLSP &NNLSP & $\tau_{NLSP}$ &$c\, \tau_{NLSP}$\\
			Points &(GeV)&(GeV)& (GeV)&(ns)& (cm)\\ 
			\hline
			\hline
				BP5  & 153.10  & 174.825 & 174.830 & 0.092  & 2.75\\
			\hline
			BP6 & 484.05 & 499.947 & 499.952 & 1.23 & 36.70\\
			\hline
			BP7 & 330.80 & 348.56 & 348.57 & 4.482 & 134.18\\
			\hline
			\hline
		\end{tabular}
		\caption{Benchmark points selected from scenario Sc 2 for collider study consistent with the $\sim 125$ GeV Higgs mass where the lifetime of NLSP is given as $\tau_{NLSP}$ and the proper decay length of NLSP is given as $c\,\tau_{NLSP}$.  This scenario has a triplino like NLSP $\tilde{ \chi}^0_2$ and nearly degenerate chargino-like NNLSP $\tilde{ \chi}^{\pm}_1$ .}\label{bps2}
	\end{center}
\end{table}

\begin{table}
	\begin{center}
		\renewcommand{\arraystretch}{1.4}
		\begin{tabular}{||c|c|c|c|c|c|c|c||}
			\hline\hline
			Benchmark&$\sigma_{\tilde{\chi}_1^{\pm}\tilde{\chi}_1^{\mp}}$ &$\sigma_{\tilde{\chi}_1^{\pm}\tilde{\chi}_1^{0}}$& $\sigma_{\tilde{\chi}_1^{\pm}\tilde{\chi}_2^{0}}$&$\sigma_{\tilde{\chi}_1^{\pm}\tilde{\chi}_2^{\mp}}$ &$\sigma_{\tilde{\chi}_2^{0}\tilde{\chi}_2^{0}} $&$\sigma_{\tilde{\chi}_2^{0}\tilde{\chi}_1^{0}}$ &$\sigma_{\tilde{\chi}_2^{0}\tilde{\chi}_2^{\pm}}$\\
			Points & (fb) &(fb)&(fb)&(fb)&(fb)&(fb)&(fb) \\ \hline
			\hline
			BP5 & 1289 & 8.29$\times 10^{-2}$ &2565.7 & 2.6$\times 10^{-6}$   & $<10^{-8}$ & $<10^{-7}$  & 1.23$\times 10^{-5}$  \\
			\hline
			BP6 & 20.95 & 2.68$\times 10^{-3}$ & 44.01 & $<10^{-7}$ & $<10^{-10}$ & $<10^{-9}$ &7.83 $\times 10^{-6}$\\
			\hline
			BP7 & 94.13  & 1.16$\times 10^{-3}$  & 192.8 & $<10^{-7}$ & $<10^{-11}$ & $<10^{-9}$ & 3.5$\times 10^{-5}$\\
			\hline
			\hline
		\end{tabular}
		\caption{ Pair and associated production cross sections for $\tilde{\chi}_{1,2}^{\pm}$ and $\tilde{\chi}_{1,2}^{0}$ at 14 TeV for each benchmark point  in scenario Sc 2.  }\label{crossec2}
	\end{center}
\end{table}
Table~\ref{crossec2} presents the cross-section for the electroweak chargino and neutralino production processes for the benchmark points in scenario Sc 2 at the LHC with center of mass energy of 14 TeV. From Table~\ref{crossec2} we see that the dominant contribution to $\ntrl2$ production comes from  $\ntrl2 \chi^\pm_1$ production at the LHC.  Such neutralino NLSP will decay via off-shell $Z/\gamma$ and gives rise to the di-lepton in the final state, 
\be\label{nu2ch1}
pp \to \ntrl2 \chpm1 \to \ell^\pm \ell^\mp \ntrl1 \ell^\pm \ntrl 1.
\ee

Eq.~\ref{nu2ch1} describes the corresponding decay topology and the final state consists of three leptons and missing energy. Due to the compressed spectrum the parameter space is still allowed from tri-lepton plus 
missing energy data \cite{winoNLSPATLAS}. Here only the lepton pair coming from $\ntrl2 $ decay is displaced while the third lepton coming from $\ntrl1$ decay  is a prompt one. 

\begin{table}
	\begin{center}
		\renewcommand{\arraystretch}{1.4}
		\begin{tabular}{||c||c|c|c||}
			\hline\hline
			Decay	 &\multicolumn{3}{|c|}{Benchmark 	Points}\\
			\cline{2-4}
			Modes	&BP5&BP6&BP7\\
			\hline
			$\ntrl2 \to \nu \bar{\nu} \ntrl1$&0.11&$2.7\times 10^{-4}$&$2.0\times 10^{-4}$\\
			\hline
			$\ntrl2 \to \ell \bar{\ell} \ntrl1$&0.02&$1.4\times 10^{-4}$&$9.8\times 10^{-4}$\\
			\hline
			$\ntrl2 \to \tau \bar{\tau} \ntrl1$&0.04&$4.7\times 10^{-3}$&$3.7\times 10^{-4}$\\
			\hline
			$\ntrl2 \to q \bar{q} \ntrl1$&0.82&0.99&0.99\\
			\hline
			$\chpm1 \to \ell \bar{\nu} \ntrl1$&0.11&0.11&0.11\\
			\hline
			$\chpm1 \to \tau \bar{\nu} \ntrl1$&0.11&0.11&0.11\\
			\hline
			$\chpm1 \to q q' \ntrl1$&0.67&0.67&0.67\\
			\hline \hline
		\end{tabular}
		\caption{Branching fraction of $\ntrl2$ and $\chpm1$ for the benchmark points in scenario Sc 2.  }\label{br2}
	\end{center}
\end{table}

Table~\ref{br2} presents the decay branching fractions of $\ntrl2$ and $\chpm1$ for the benchmark points in scenario Sc 2.  It can be seen that in BP5, the $\ntrl2$ decays to charged lepton pairs by 2-4\% and for other BPs, the branching fraction is $\sim 10^{-4}$. For all the benchmark points $\ntrl2$ dominantly decays into  $b\bar{b}\ntrl1$ with branching fractions 53\%, 99\% and 99\% for BP5, BP6 and BP7, respectively,  giving rise to displaced jets.

The multi-lepton plus missing energy bounds in \cite{winoNLSPATLAS} do not exclude our parameter space due to the following reasons. First, the charged leptons coming from  $\chpm1$ or $\ntrl2$ are coming from three-body decays due to the lack of phase space. Thus they are very soft and will not appear after the basic cuts demanded in \cite{winoNLSPATLAS} for the electron and muon, which are $\geq 20$ and 30 GeV, respectively. Second, the decay branching fraction of the leptonic modes is much smaller in our case. We see from Table~\ref{br2} that the most dominant mode is the hadronic one. We also should not forget that for some benchmark points the production cross-sections are less compared to the MSSM case, where either wino- or Higgsino-like NLSPs are considered.  $\chpm1$ which is NNLSP in this only gives rise to prompt leptons and has no contribution towards displaced charged leptonic signature as discussed earlier.

\begin{figure}[tbh]
	\begin{center}
	\mbox{\subfigure[]{\includegraphics[width=0.35\linewidth, angle=-90]{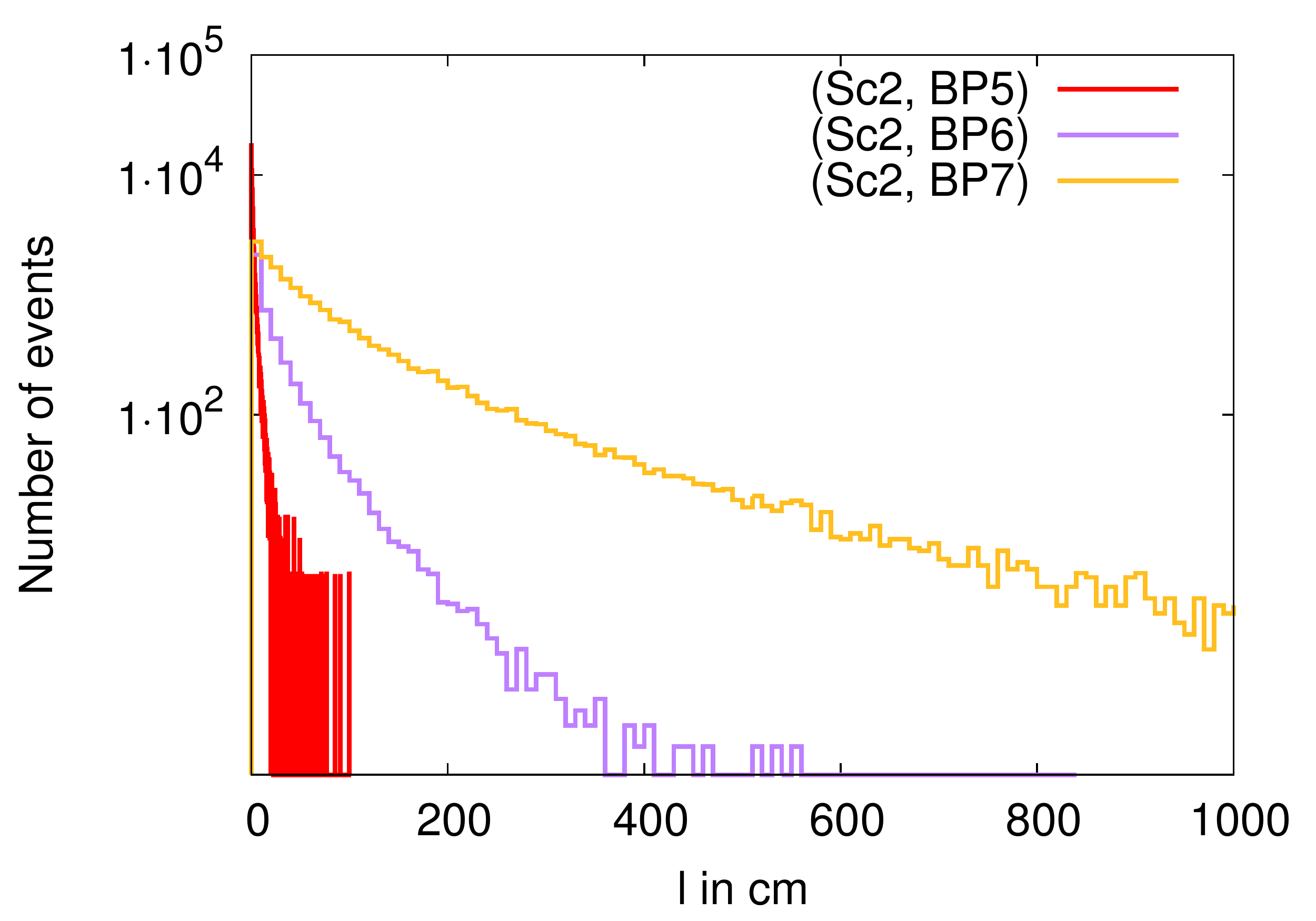}}
		\subfigure[]{\includegraphics[width=0.35\linewidth, angle=-90]{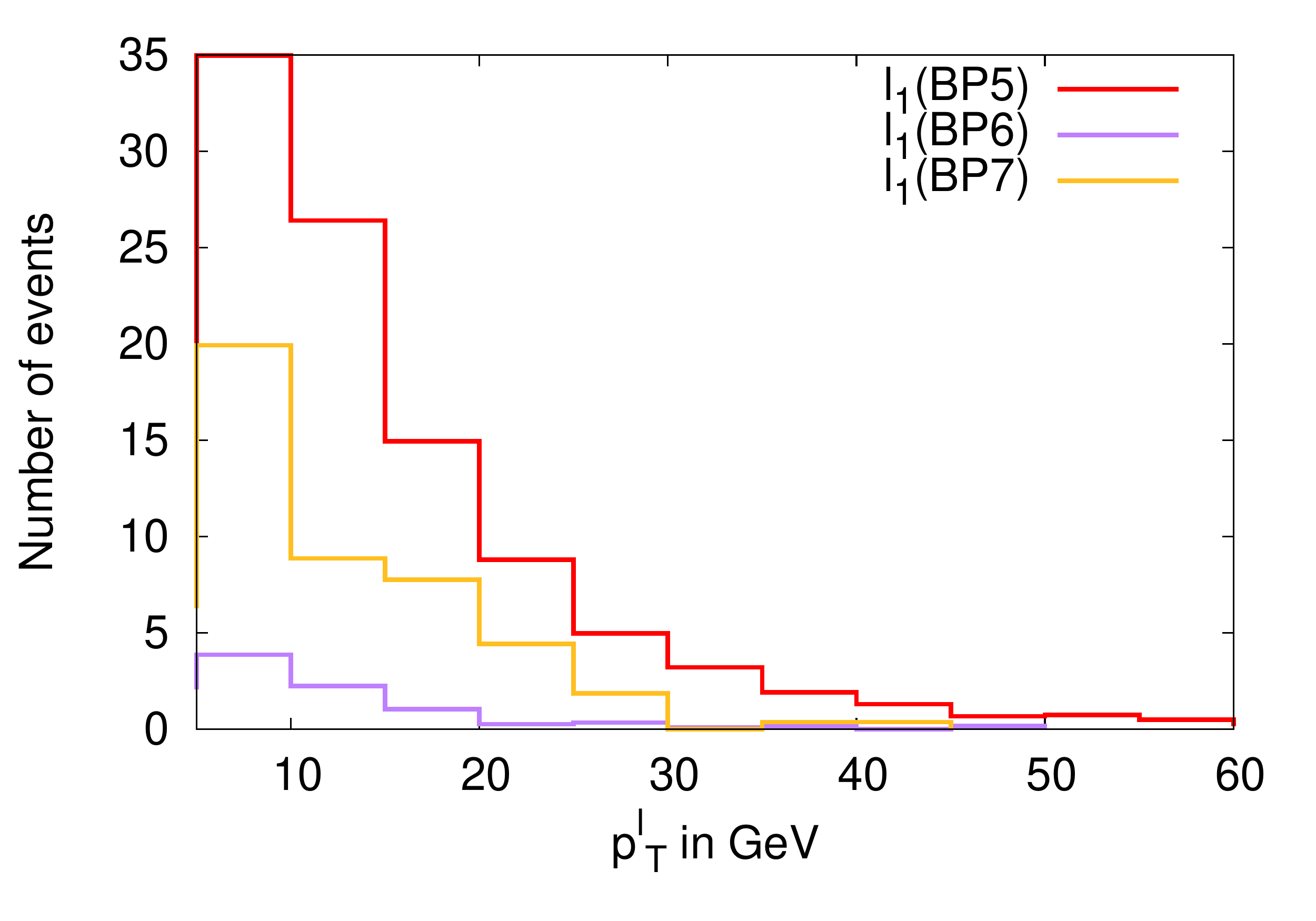}}}
		\mbox{\subfigure[]{\includegraphics[width=0.35\linewidth, angle=-90]{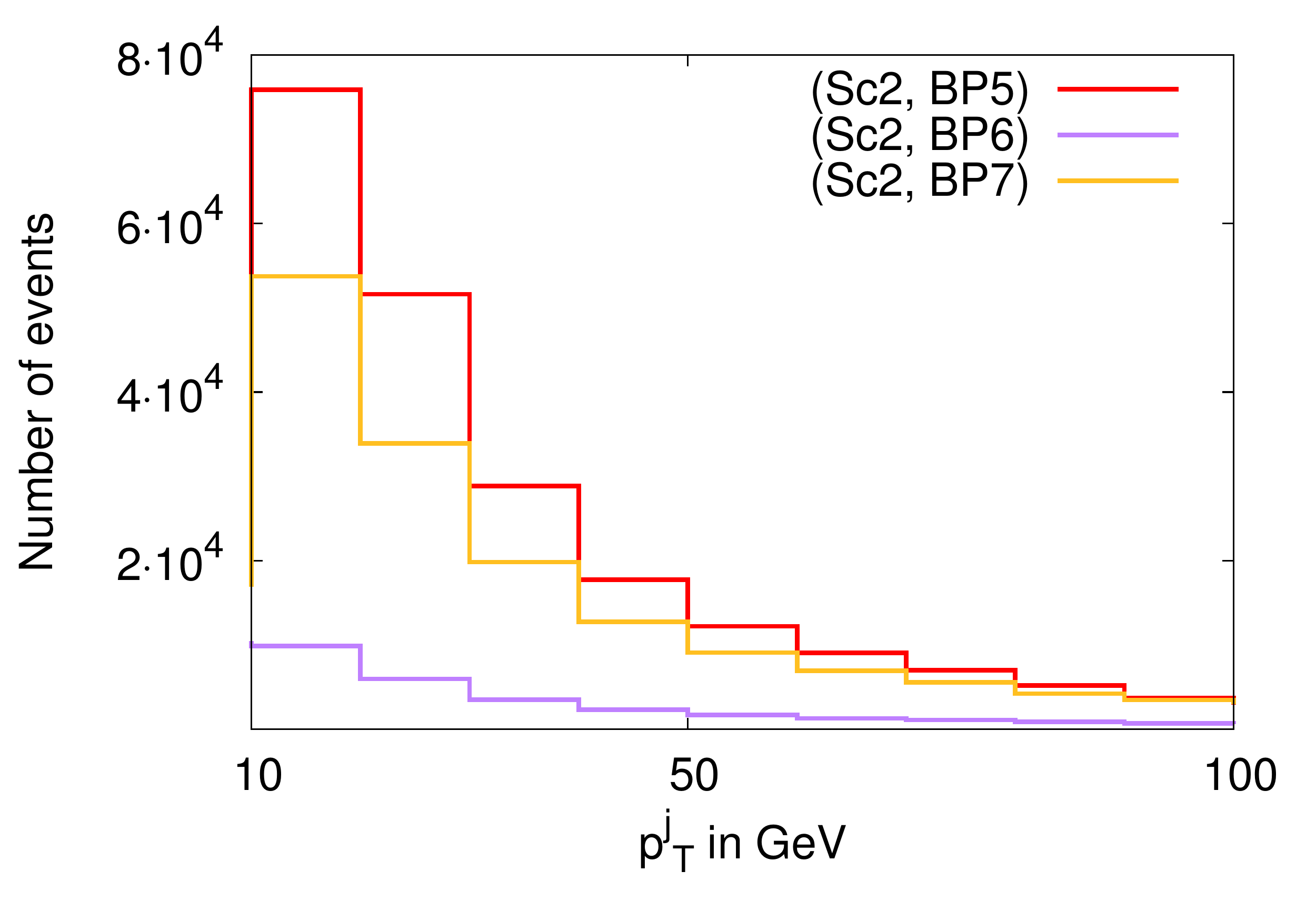}}}
	\caption{(a)Transverse displaced decay length distribution, (b) transverse charged lepton momentum distribution and (c) transverse jet $p_T$ momentum distribution for the benchmark points at the LHC with 14 TeV.}\label{dissc2}
\end{center}
\end{figure}


\begin{table}
	\begin{center}
		\renewcommand{\arraystretch}{1.2}
		\begin{tabular}{||c|c|c|c||}
			\hline\hline
			Benchmark Points&\multicolumn{1}{|c|}{$n_{\ell}\geq 1$} &\multicolumn{1}{|c|}{$n_{\ell}\geq 2$}& \multicolumn{1}{|c|}{$n_{\ell}\geq 3$}\\
			\hline
			BP5 &7799.7$\begin{cases}
			3586.8\\
			3940.9\\    
			272.0	\\
			0.0   
			\end{cases}$&  4038.4& 543.9\\
			\hline
			BP6 &2.4$\begin{cases}
			0.1\\
			1.0\\    
			1.1\\
			0.2    
			\end{cases}$&1.0&0.0\\
			\hline
			BP7&3.8$\begin{cases}
			0.4\\
			1.5\\    
			1.9\\
		    0.0   
			\end{cases}$&1.2	&0.0\\
			\hline
			\hline
		\end{tabular}
		\caption{ Number of events with multiple charged leptons with at least one of them is displaced with displacement  0.1mm  to 10 m at 14 TeV for the benchmark points in scenario 2.  }\label{sc22}
	\end{center}
\end{table}


\begin{table}
	\begin{center}
		\renewcommand{\arraystretch}{1.4}
		\begin{tabular}{||c|c|c|c||}
			\hline\hline
			Benchmark&$n_{j}\geq 1$ &$n_{j}\geq 2$& $n_{b}\geq 1$\\
			Points & &&  \\ \hline\hline
			BP5 &161818.7$\begin{cases}
			67098.2\\
			88285.7\\    
			6414.3\\
			20.5
			\end{cases}$&82595.1$\begin{cases}
			31178.4\\
			 46895.9\\    
			  4505.4\\
			15.4
			\end{cases}$
			&389.9$\begin{cases}118.0\\
			236.0\\
			  35.9\\
			 0.0\end{cases}$\\
			\hline
			BP6 &2783.0$\begin{cases}
			163.7\\
			 924.7\\    
			1570.1\\
			124.5
			\end{cases}$&1448.7$\begin{cases}
			 81.3\\
			 460.0\\    
			829.3\\
	     	78.1
			\end{cases}$ &5.3$\begin{cases}
			0.0\\
			1.5\\    
			3.0\\
			0.8
			\end{cases}$  \\
			\hline
			BP7&12251.6.1$\begin{cases}
			 197.0\\
			1458.7\\    
			6471.9\\
			 4124.0
			\end{cases}$&6412.8$\begin{cases}
			92.9	\\
			706.4\\
	      3271.8\\
	        2341.7\end{cases}$
			&24.3$\begin{cases}0.0\\
			1.9\\
			10.4\\
			12.0\end{cases}$\\
			\hline
		\end{tabular}
		\caption{ Number of events with multiple jets with at least one of them is displaced with displacement  0.1mm to 10 m at 14 TeV for the benchmark points in scenario Sc 2.  }\label{sc23}
	\end{center}
\end{table}
The jets and charged leptons coming from the displaced three-body decays of $\ntrl2$ are very soft. Fig.~\ref{dissc2} (a) shows the displaced transverse decay lengths for the benchmark points  at the LHC with center of mass energy of 14 TeV. We see that BP5 can have a few cm of displaced decay whereas BP6 and BP7 can go up to few meters. The charged lepton $p_T$ distribution can be seen from Figure~\ref{dissc2} (b) and it is evident that leptons are rather soft.
Similarly Fig.~\ref{dissc2} (c) presents the $p^j_T$ distributions for the benchmark points and the jets are rather soft. The compressed spectrum thus prompts us to choose rather soft $p_T$ cuts for leptons ($p^\ell_T\geq 5$ GeV) and jets ($p^\ell_T\geq 10$ GeV). 

We choose events where we have at least one displaced charged lepton with different displaced decay lengths for the multi-lepton final states for the benchmark points. Such events 
at the LHC with 14 TeV center of mass energy at 100 fb$^{-1}$ integrated luminosity  are collected in Table~\ref{sc22}. For single displaced lepton events we again decompose the displaced length $\ell_d$ in four different ranges: $0.1 \,\text{mm}\, <\ell_d \leq1$ cm, $1 \,\text{cm}\, <\ell_d \leq10$ cm, $10 \,\text{cm}\, <\ell_d \leq1$ m and $1 \,\text{m}\, <\ell_d \leq10$ m, respectively. The presence of one displaced lepton makes the final states completely background free. We see that only BP5 has promising number of events at 100 fb$^{-1}$ of integrated luminosity, for other benchmark points one has to wait for high-luminosity (HL), i.e., $\mathcal{O}(3000)$ fb$^{-1}$  at the LHC. Only BP5 gives rise to sufficient number of tri-lepton events at 100 fb$^{-1}$ of integrated luminosity. 

Table~\ref{sc23} presents the numbers of events at the LHC with center of mass energy of  14 TeV and at an luminosity of 100 fb$^{-1}$ with multiple jets produced via a displaced decay. The third column lists such events with at least one displaced $b$-jet. The soft jets have comparatively lower $b$-tagging efficiency \cite{btag}, nevertheless due to large branching ratio of $\ntrl2$ into $b\bar{b}\ntrl1$ this final state looks promising. It is interesting to see that only $\ntrl2$ contributes to the final states with $b$-jets via its decay to off-shell $Z$ or Higgs boson. On the other hand,  the lightest chargino, which is NNLSP in this scenario, mainly decays via off-shell $W^\pm$, and thus does not contribute to $b$-jet final states. A truly tagged $b$-jet is a displaced one as it comes from the $\ntrl2$ decay. 

\subsection{Sc 3: Higgsino LSP}
Similar to Sc1, Sc 3  has a chargino NLSP, which is nearly degenerate with the LSP but unlike Sc1, here it is a Higgsino-like LSP. Table~\ref{bpsc3} presents the benchmark points for scenario Sc 3. Table~\ref{crossec3} l
ist the cross-sections for the benchmark points at the LHC with center of mass energy of 14 TeV. Due to very small mass gap the decay products, mostly the charged leptons, cannot cross the threshold $p_T$ cuts, giving 
disappearing charged track as a signal. In this case the benchmark points (BP8, BP9 and BP10) are having NLSP with displaced decay lengths mm to cm, much smaller than most of the benchmark points in other 
scenarios. However, we will see that due to some hard ISR jets the final state decay products carry momenta above the threshold and such decay products i.e., the charged leptons and jets can be detected.

\begin{table}
	\begin{center}
		\renewcommand{\arraystretch}{1.4}
		\begin{tabular}{||c|c|c|c|c|c|c||}
			\hline\hline
			Benchmark&NLSP&LSP mass &NLSP mass &NNLSP mass & $\tau_{NLSP}$ &$c\, \tau_{NLSP}$\\
			Points & &(GeV)&(GeV)& (GeV)&(ns)& (cm)\\ \hline\hline
			BP8 & $\tilde{\chi}_1^{\pm}$ & 113.648 & 114.476 & 195.124 & 0.0038 & 0.113 \\
			\hline
			BP9 & $\tilde{\chi}_1^{\pm}$ & 367.33 & 368.161 & 439.22 & 0.0028 & 0.082\\
			\hline
			BP10 & $\tilde{\chi}_1^{\pm}$  & 177.38 & 177.73 & 211.401 & 0.274 & 8.16\\
			\hline
			
		\end{tabular}
		\caption{Benchmark points for a collider study consistent with the $\sim 125$ GeV Higgs mass where the lifetime of NLSP is given as $
			\tau_{NLSP}$ and the proper decay length of NLSP is given as $c\,\tau_{NLSP}$.   }\label{bpsc3}
	\end{center}
\end{table}
\begin{table}
	\begin{center}
		\renewcommand{\arraystretch}{1.4}
		\begin{tabular}{||c|c|c|c|c|c|c|c||}
			\hline\hline
		Benchmark&$\sigma_{\tilde{\chi}_1^{\pm}\tilde{\chi}_1^{\mp}}$ &$\sigma_{\tilde{\chi}_1^{\pm}\tilde{\chi}_1^{0}}$& $\sigma_{\tilde{\chi}_1^{\pm}\tilde{\chi}_2^{0}}$&$\sigma_{\tilde{\chi}_1^{\pm}\tilde{\chi}_2^{\mp}}$ &$\sigma_{\tilde{\chi}_2^{0}\tilde{\chi}_2^{0}} $&$\sigma_{\tilde{\chi}_2^{0}\tilde{\chi}_1^{0}}$ &$\sigma_{\tilde{\chi}_2^{0}\tilde{\chi}_2^{\pm}}$\\
			Points & (fb) &(fb)&(fb)&(fb)&(fb)&(fb)&(fb) \\ \hline
			\hline
			BP8 & 3495.03 & 275.86 & 547.89  & 62.30 & 1.11 & 86.12  & 99.48\\
			\hline
			BP9 & 44.52 & 3.70 & 13.40 & 4.31 & 1.4$\times 10^{-3}$  &  2.14 & 6.93\\
			\hline
			BP10 &693.80  &  53.08 & 219.04 & 78.70  & 0.289  & 41.43 & 115.1 \\
			\hline
		\end{tabular}
		\caption{ Pair and associated production cross sections for $\tilde{\chi}_{1,2}^{\pm}$ and $\tilde{\chi}_{1,2}^{0}$ at 14 TeV for each benchmark point  in scenario 3.  }\label{crossec3}
	\end{center}
\end{table}

\begin{figure}[tbh]
	\begin{center}
	\mbox{\subfigure[]{	\includegraphics[width=0.33\linewidth, angle=-90]{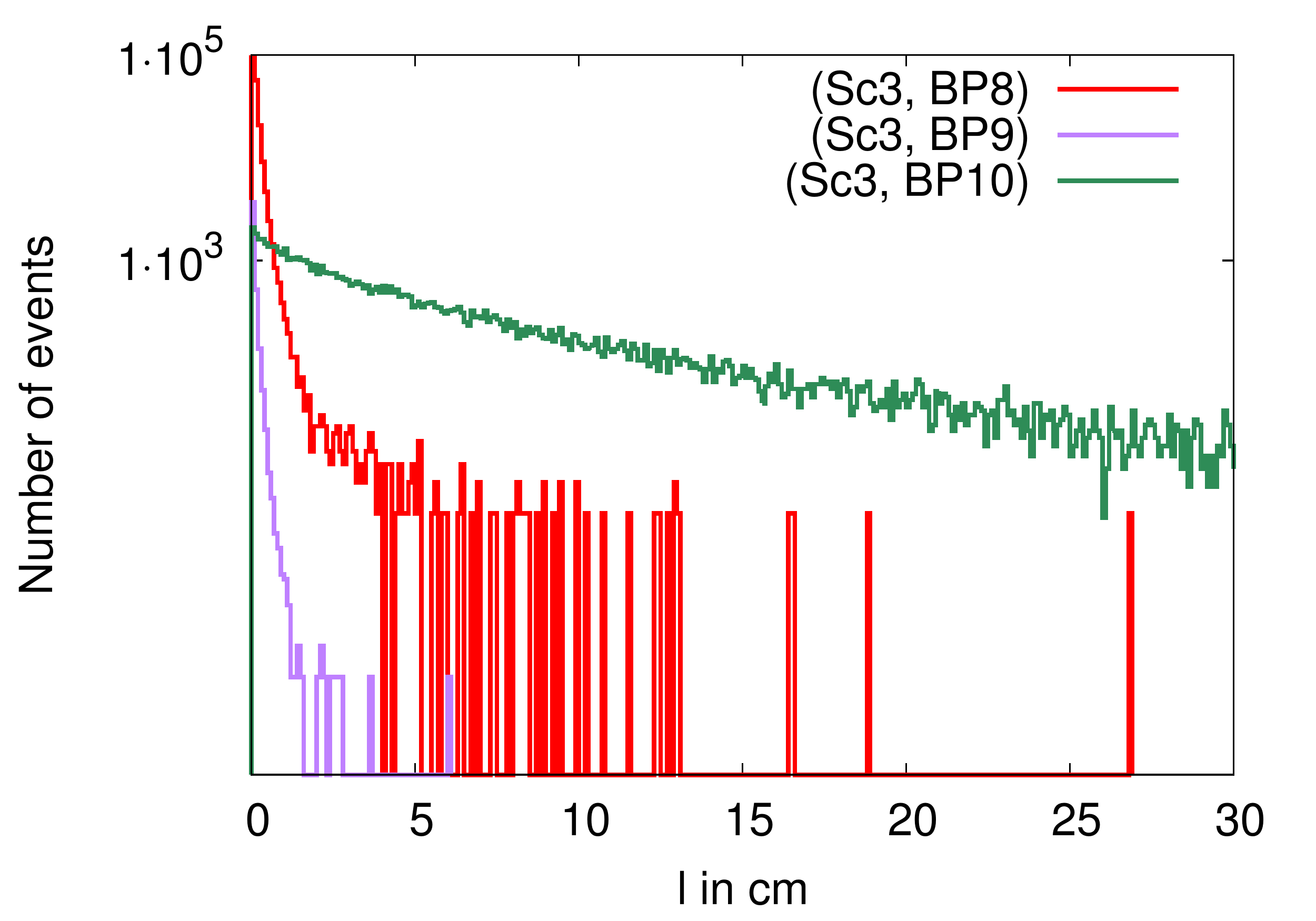}}
		\subfigure[]{	\includegraphics[width=0.33\linewidth, angle=-90]{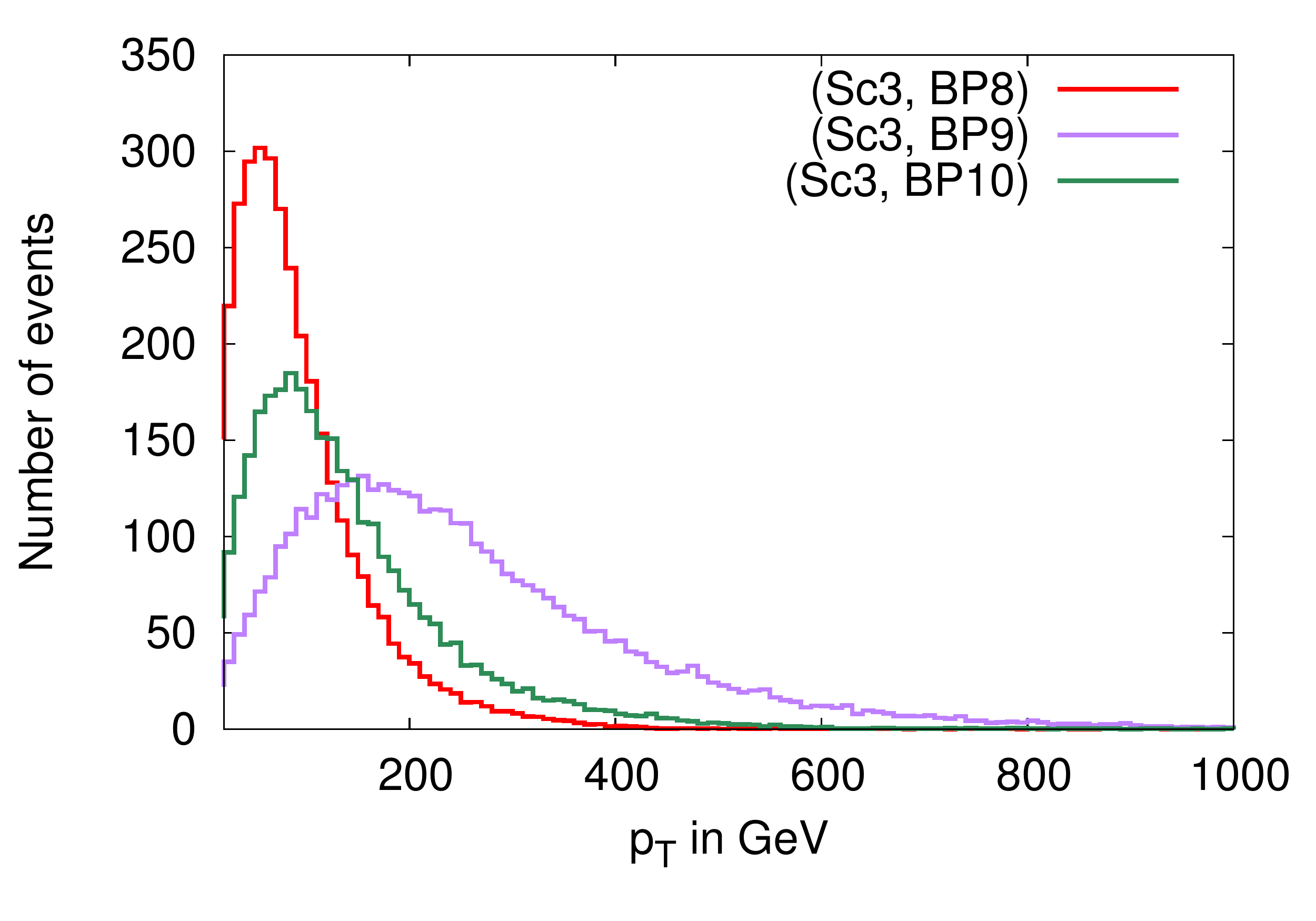}}}
		\caption{The distribution for (a) the transverse displaced decay length  and (b) the $p_T$ of the chargino NLSP for the benchmark points in Sc3
			at the LHC with 14 TeV.}\label{dplsc3}
	\end{center}
\end{figure}

\begin{table}[t]
	\begin{center}
		\renewcommand{\arraystretch}{1.2}
		\begin{tabular}{||c|c|c|c|c||}
			\hline\hline
			Benchmark 	Points &Process &BP8&BP9& BP10\\
			\hline\hline
			\multirow{3}{*}{0.1-1cm} &$\tilde{\chi}_1^{\pm}\tilde{\chi}_1^{\mp}$&290786.5&3441.6&14816.8\\
			&$\tilde{\chi}_1^{\pm}\tilde{\chi}_1^{0}$&23810.6& 304.7&883.1\\
			&$\tilde{\chi}_1^{\pm}\tilde{\chi}_2^{0}$& 47382.5&1057.3&4340.5\\
			\hline
			\multirow{3}{*}{1-10cm} &$\chi_1^{\pm}\tilde{\chi}_1^{\mp}$& 1481.9&2.6&41067.4\\
			&$\tilde{\chi}_1^{\pm}\tilde{\chi}_1^{0}$& 163.3& 0.3&3054.9\\
			&$\tilde{\chi}_1^{\pm}\tilde{\chi}_2^{0}$&373.6&1.3&12902.8\\
			\hline
			\multirow{3}{*}{10cm -1m}&$\tilde{\chi}_1^{\pm}\tilde{\chi}_1^{\mp}$&69.9&0.0&13221.1\\
			&$\tilde{\chi}_1^{\pm}\tilde{\chi}_1^{0}$&0.6&0.0&1418.4\\
			&$\tilde{\chi}_1^{\pm}\tilde{\chi}_2^{0}$& 2.2&0.0&4576.2\\
			\hline
			\multirow{3}{*}{1-10m }&$\tilde{\chi}_1^{\pm}\tilde{\chi}_1^{\mp}$& 0.0&0.0& 63.8\\
			&$\tilde{\chi}_1^{\pm}\tilde{\chi}_1^{0}$&0.0&0.0&11.0\\
			&$\tilde{\chi}_1^{\pm}\tilde{\chi}_2^{0}$&0.0&0.0& 30.2\\
			\hline
			\hline
		\end{tabular}
		\caption{ Number of displaced NLSP decays which can have charged track in the ranges of $0.1-1$ cm, $1-10$ cm, $0.1-1$ m and $1-10$ m for the benchmark points of scenario Sc 3 at the LHC with 14 TeV center of mass energy and at an integrated luminosity of 100 fb$^{-1}$.  }\label{sc3}
	\end{center}
\end{table}
In Fig.~\ref{dplsc3} (a) we show the transverse decay lengths of the the lightest chargino NLSP for the benchmark points in this scenario. It can be seen that the displaced transverse length can go up to few tens of cm for 
BP8 and BP9, and for BP10 the reach could be in meter range.  Fig.~\ref{dplsc3} (b) shows the momentum that is carried by the lightest chargino NLSP for the benchmark points in this scenario, which could be very hard, specially for BP9. Although  this is nearly degenerate NLSP-LSP  scenario, the decay product can have $\mathcal{O}$(10) GeV momentum from the choice of phase-spaces with higher momenta.

Table~\ref{sc3} gives numbers of events with displaced lightest chargino NLSP decay for the range of 0.1 mm to 10 m for the benchmark points. As anticipated only BP10 has events with $\mathcal{O}(10)$ meter of displacements.  Here we have considered the dominant contributions from $\chpm1\chmp1,\, \chpm1\ntrl1,\, \chpm1\ntrl2$ production processes. 
$\ntrl2$ is the NNLSP and has a prompt decay for all three benchmark points. However, it can decay to $ \chpm1$ which can give rise to additional displaced decays. Out of these events most of the events will end up giving disappearing charged track as the decay products will be below the initial trigger cuts. Nevertheless, we will bank on the possibility of the boosted decay events where the NLSP with higher momentum forward some momentum to the decay products i.e., the charged leptons and jets. The other possibility is that the final state gets high momentum recoil  due to some ISR jets. Both these effects are incorporated in our PYTHIA based analysis. 

\begin{table}
	\begin{center}
		\renewcommand{\arraystretch}{1.4}
		\begin{tabular}{||c|c|c|c|c|c|c||}
			\hline\hline
			Benchmark&\multicolumn{3}{|c|}{$n_{\ell}\geq 1$} &\multicolumn{3}{|c|}{$n_{\ell}\geq 2$}\\
			\hline
			Points & $\chpm1\chmp1$&$\chpm1\ntrl1$&$\chpm1\ntrl2$&$\chpm1\chmp1$&$\chpm1\ntrl1$&$\chpm1\ntrl2$\\ \hline\hline
			BP8 & 636.1$\begin{cases}
			223.7\\
		    349.5\\    
			 62.9\\
			0.0    
			\end{cases}$&18.3	$\begin{cases}
				10.4\\
				 8.3\\    
				0.6\\
				0.0    
			\end{cases}$&29.9$\begin{cases}
			 17.4\\
			12.5\\    
			0.0\\
			0.0    
			\end{cases}$&118.8&0.0&5.0\\
			\hline
			BP9 &3.3$\begin{cases}
			2.3\\
			 1.0\\    
			0.0\\
			0.0 
			\end{cases}$ &0.1$\begin{cases}
			0.1\\
			0.0\\    
			0.0\\
			0.0 
			\end{cases}$ &0.4$\begin{cases}
			0.3\\
			0.1\\    
			0.0\\
			0.0    
			\end{cases}$&0.0&0.0&0.1\\
			\hline
			BP10&29.2$\begin{cases}
			0.0\\
			2.8\\    
		    9.7\\
			16.7 
			\end{cases}$ &1.4$\begin{cases}
			0.0\\
			0.0\\    
			0.7\\
			0.7 
			\end{cases}$ &4.8$\begin{cases}
			0.0\\
			0.0\\    
			 2.2\\
		   2.6
			\end{cases}$ &2.8&0.0&0.9\\
			\hline
			\hline
		\end{tabular}
		\caption{ Number of events with multiple charged leptons with at least one of them displaced with displacement  0.1 mm  to 10 m at 14 TeV for the benchmark points in scenario Sc 3. Here the leptons are rather soft $p^{\ell}_T \geq 5$ GeV and at least one of them is displaced. }\label{sc32}
	\end{center}
\end{table}

Table~\ref{sc32} presents the single and di-lepton ($e,\,\mu$) final state numbers for the benchmark points at the LHC with center of mass energy of 14 TeV and at an integrated luminosity of 100 fb$^{-1}$, where at least one of the lepton is having displaced production. However, it is important to remember that we have put the minimum $p_T$ cut of 5 GeV for these leptons as they are very soft due to nearly degenerate scenario.  A demand of much higher momentum will push us to high $p_T$ corner of the phase space and we will loose in event numbers for the final states. It is evident that finding one displaced lepton could be possible but for higher lepton multiplicity such events are rare. We see BP8 as a good possibility for di-lepton events but the tri-lepton final state looks very illusive.


\begin{table}
		\begin{center}
		\renewcommand{\arraystretch}{1.4}
		\begin{tabular}{||c|c|c|c||}
			\hline\hline
			Benchmark&\multicolumn{3}{|c|}{$n_{j}\geq 1$} \\
			\hline
			Points&$\chpm1\chmp1$&$\chpm1\ntrl1$&$\chpm1\ntrl2$\\ \hline\hline
			BP8 & 217237.1$\begin{cases}
			 215943.9\\
			1286.2\\    
			7.0\\
			0.0
			\end{cases}$&18028.6$\begin{cases}
			   17898.9\\
		 129.7\\    
			0.0\\
			0.0
			\end{cases}$&44478.7$\begin{cases}
			44117.6\\
			361.1\\    
			0.0\\
			0.0
			\end{cases}$\\
			\hline
			BP9 &2791.7$\begin{cases}
			 2789.7\\
			2.0\\    
			0.0\\
			0.0	
			\end{cases}$&248.0$\begin{cases}
			247.8\\
			0.2\\    
			0.0\\
			0.0
			\end{cases}$&1010.1$\begin{cases}
			1009.1\\
			1.0\\    
			0.0\\
			0.0
			\end{cases}$\\
			\hline
			BP10&53128.3$\begin{cases}
			11229.8\\
			31501.3\\    
		   10357.0\\
			40.2
			\end{cases}$&4183.7$\begin{cases}
			683.5\\
			2372.6\\    
			1119.8\\
	     	7.8
			\end{cases}$&18826.9$\begin{cases}
			3691.7\\
			11089.6\\    
			4023.3\\
			 22.3
			\end{cases}$\\
			\hline
		\end{tabular}
		\caption{ Number of events with multiple jets with at least one of them displaced with displacement  0.1 mm to 10 m at 14 TeV for the benchmark points in scenario Sc 3.  }\label{sc31j}
	\end{center}
\end{table}
Next we study the status of the displaced jet final states. Table~\ref{sc31j} shows the numbers of events with at least one displaced jet with $p^j_T \geq 10$ GeV for the benchmark points in this scenario at the LHC with center of mass energy of 14 TeV with an integrated luminosity of 100 fb$^{-1}$. We see that as expected, only BP10 has some displaced jet events in the meter range. Due to the degenerate NLSP-LSP case, we demanded much lower momentum cuts on jets. A requirement of $p^j_T \geq 20$ GeV reduces the event number in the final states by $43\%$.
\begin{table}
		\begin{center}
	\renewcommand{\arraystretch}{1.4}
	\begin{tabular}{||c|c|c|c||c||}
		\hline\hline
		Benchmark&\multicolumn{3}{|c|}{$n_{j}\geq 2$}& $n_{b}\geq 1$\\
		\cline{2-5}
		Points&$\chpm1\chmp1$&$\chpm1\ntrl1$&$\chpm1\ntrl2$&$\chpm1\ntrl2$\\ \hline\hline
		BP8 &15516.9$\begin{cases}
		149692.1\\
		817.8\\    
		7.0\\
		0.0
		\end{cases}$
		& 12670.2$\begin{cases}
		12580.3\\
		89.9\\    
		0.0\\
		0.0
		\end{cases}$&38549.1$\begin{cases}
		38260.2\\
		 288.9\\    
		0.0\\
		0.0
		\end{cases}$& 580.3$\begin{cases} 575.3\\
		5.0\\
		0.0\\
		0.0\end{cases}$\\
		\hline
		BP9 &2046.0$\begin{cases}
		2045.1\\
		0.9\\    
		0.0\\
		0.0
		\end{cases}$ &182.6$\begin{cases}
		182.4\\
		0.2\\    
		0.0\\
		0.0
		\end{cases}$ &889.9$\begin{cases}
		888.9\\
		1.0\\    
		0.0\\
		0.0
		\end{cases}$&  29.6$\begin{cases}
		29.6 \\
		0.0\\    
		0.0\\
		0.0
	\end{cases}$ \\
		\hline
		BP10&37861.9$\begin{cases} 7906.5\\
		22363.9\\
		7570.7\\
		20.8	\end{cases}$
		&3022.3$\begin{cases} 493.7\\
		1706.2\\
		 818.0\\
		4.4
			\end{cases}$&14462.3$\begin{cases} 2791.0\\
		8524.2\\
		3128.3\\
		18.8	\end{cases}$&47.3$\begin{cases}
		7.9\\
		25.0\\
		14.0\\
		0.4\end{cases}$\\
		\hline
	\end{tabular}
	\caption{ Number of events with di-jet with at least one of them displaced with displacement  0.1 mm to 10 m at 14 TeV for the benchmark points in scenario Sc 3. The last column shows if at least one of them is a displaced $b$-jet. }\label{sc32j}
		\end{center}
\end{table}

Table~\ref{sc32j} shows the number of events for the di-jet final states for the benchmark points at the LHC with 14 TeV of center of mass energy at an integrated luminosity of 100 fb$^{-1}$, where we demand to have at least one of the jets to be produced via displaced decay of the NLSP.
The dominant contributions are from $\chpm1\chmp1, \, \chpm1\ntrl1,\, \chpm1\ntrl2$, respectively. The requirement of soft jets are again motivated from the compressed mass spectrum for NLSP-LSP and a demand of 
$p^j_T \geq 20$ GeV will reduce the number of events by 58-65\%. In that we need to go for higher luminosity LHC in order of have sufficient number of events.

Finally in the last column of Table~\ref{sc32j} we present the number of events where we have a least one displaced $b$-jet in the final state. Such $b$-jets produced via the displaced decay of NLSP can be really promising. Along with the displaced charged leptons it can give additional handle for the system.  

\section{Discussion and conclusion}\label{concl}
In this article we have considered the phenomenology of the electroweak gaugino sector for the $Y=0$ triplet extended supersymmetric SM. The triplet extension is motivated due to reducing the demand for large SUSY mass scale for a desired $\sim 125$ GeV Higgs boson. Such extensions specially with $Y=0$ do not couple to fermions and give rise to 
interesting phenomenology in the neutral and charged Higgs sectors \cite{PBAS, PBAS2,PBAS3, TNSSM1, TNSSMch}.

Similar to Higgs sectors the phenomenology of electroweak chargino and neutralino sectors differ from MSSM and is thus worth exploring. 
We noticed that triplet-like charginos and neutralinos are almost mass degenerate. In scenario Sc 1, such triplino-like low lying states give rise to displaced phenomenology. In scenario Sc 2, both the NLSP and NNLSP are of triplino-type and they are also nearly degenerate  as mass eigenstates tend to follow the same gauge structure. Similar feature for Higgs mass eigenstates following the gauge structure in a supersymmetric extended Higgs scenario has already been observed \cite{TNSSM1, TNSSMch}. The triplino-like chargino and triplet charged Higgs boson do not couple to the fermions and alter the bounds coming from the rare decays, such as $B \to X_s \gamma$ \cite{PBAS2}.

Unlike scalar component of the $Y=0$ fermionic triplet in supersymmetric Type-III seesaw \cite{ejc1, ejc2}, these triplinos are fermions, do not carry any lepton numbers and couple to Higgs boson via Type-III Yukawa coupling. Type-III model scalar triplino decay can give rise to a doublet-type  (viz. $h_{125}$) displaced Higgs production \cite{ejc2}. Such features can be explored in order to distinguish the $Y=0$ $SU(2)$ triplets with different spins. Generically seesaw models predict displaced decays due to very small Yukawa couplings  \cite{seesaw}. 

Displaced jets  can come from various other models including $R$-parity violating decays 
and recent LHC searches have put some bounds on models \cite{dispjets}. $R$-parity violating Higgs decays can also lead to displaced multi-lepton final states \cite{unusig}. In a supersymmetric  $U(1)$ extended scenario, superpartners of right-handed neutrinos can have displaced decays due to a very suppressed coupling occurring because of the cancellation among the parameters in the superpotential and the soft parameters \cite{RHNs}. However, in these cases the corresponding decay products have relatively large momentum, enough to be detected. In this study we have used rather soft triggers, i.e., $p^{\ell}_T \geq 5$ GeV and  $p^{j}_T \geq 10$ GeV. Application of larger momentum cuts i.e., $p^{\ell}_T \geq 20$ GeV and  $p^{j}_T \geq 20$ GeV reduce the leptonic and jet final state events by  $\sim 41\%$ and $\sim 43\%$, respectively.  For such large momentum cuts, the high luminosity version of LHC is essential. \section*{Acknowledgments}
PB acknowledges University of Helsinki for the visit during initial part of the project. 
PB also thanks Prof. Rahul Sinha for arranging the visit at Institute of Mathematical Sciences, Chennai for the final part of the project. ASK acknowledges Nam{\i}k Kemal University, Physics Department for the hospitality during the initial part of the project.  
KH acknowledges H2020-MSCA-RICE-2014 grant no.  645722 (NonMinimalHiggs).


\begin{thebibliography}{99}

\bibitem{Higgsd1}
S.~Chatrchyan {\it et al.}  [CMS Collaboration],
Phys.\ Lett.\ B {\bf 716} (2012) 30
[arXiv:1207.7235 [hep-ex]].

\bibitem{Higgsd2}
G.~Aad {\it et al.}  [ATLAS Collaboration],
Phys.\ Lett.\ B {\bf 716} (2012) 1
[arXiv:1207.7214 [hep-ex]].

\bibitem{carena}
M.~Carena, S.~Gori, N.~R.~Shah and C.~E.~M.~Wagner,
JHEP {\bf 1203} (2012) 014
doi:10.1007/JHEP03(2012)014
[arXiv:1112.3336 [hep-ph]].
M.~W.~Cahill-Rowley, J.~L.~Hewett, A.~Ismail and T.~G.~Rizzo,
Phys.\ Rev.\ D {\bf 86} (2012) 075015
doi:10.1103/PhysRevD.86.075015
[arXiv:1206.5800 [hep-ph]].


\bibitem{PBAS}
P.~Bandyopadhyay, K.~Huitu and A.~Sabanci,
JHEP {\bf 1310} (2013) 091
doi:10.1007/JHEP10(2013)091
[arXiv:1306.4530 [hep-ph]]
\bibitem{NMSSM}
U.~Ellwanger, C.~Hugonie and A.~M.~Teixeira,
Phys.\ Rept.\  {\bf 496} (2010) 1
doi:10.1016/j.physrep.2010.07.001
[arXiv:0910.1785 [hep-ph]].


\bibitem{epqr1}
J.~R.~Espinosa and M.~Quiros,
Nucl.\ Phys.\ B {\bf 384} (1992) 113.
doi:10.1016/0550-3213(92)90464-M
\bibitem{epqr2}
J.~R.~Espinosa and M.~Quiros,
Phys.\ Lett.\ B {\bf 279} (1992) 92.
doi:10.1016/0370-2693(92)91846-2
\bibitem{PBAS2} 
P. ~Bandyopadhyay, S.~Di Chiara, K.~Huitu and A.~S.~Keceli,
JHEP {\bf 1411}, 062 (2014)
doi:10.1007/JHEP11(2014)062
[arXiv:1407.4836 [hep-ph]].



\bibitem{PBAS3} 
P.~Bandyopadhyay, K.~Huitu and A.~Sabanci Keceli,
JHEP {\bf 1505}, 026 (2015)
doi:10.1007/JHEP05(2015)026
[arXiv:1412.7359 [hep-ph]].
\bibitem{agashe}
K.~Agashe, A.~Azatov, A.~Katz and D.~Kim,
Phys.\ Rev.\ D {\bf 84} (2011) 115024
doi:10.1103/PhysRevD.84.115024
[arXiv:1109.2842 [hep-ph]].



\bibitem{quiros}
L.~Cort, M.~Garcia and M.~Quiros,
Phys.\ Rev.\ D {\bf 88} (2013) no.7,  075010
doi:10.1103/PhysRevD.88.075010
[arXiv:1308.4025 [hep-ph]].
\bibitem{GM}
H.~Georgi and M.~Machacek,
Nucl.\ Phys.\ B {\bf 262} (1985) 463.
doi:10.1016/0550-3213(85)90325-6
\bibitem{TNSSM1}
P.~Bandyopadhyay, C.~Corian\`o and A.~Costantini,
JHEP {\bf 1509} (2015) 045
[arXiv:1506.03634 [hep-ph]].
P.~Bandyopadhyay, C.~Corian\`o and A.~Costantini,
JHEP {\bf 1512} (2015) 127
doi:10.1007/JHEP12(2015)127
[arXiv:1510.06309 [hep-ph]].

\bibitem{TNSSMch}
P.~Bandyopadhyay, C.~Corian\`o and A.~Costantini,
Phys.\ Rev.\ D {\bf 94} (2016) no.5,  055030
doi:10.1103/PhysRevD.94.055030
[arXiv:1512.08651 [hep-ph]].
P.~Bandyopadhyay and A.~Costantini,
JHEP {\bf 1801} (2018) 067
doi:10.1007/JHEP01(2018)067
[arXiv:1710.03110 [hep-ph]].

\bibitem{disch}
G.~Aad {\it et al.} [ATLAS Collaboration],
Phys.\ Rev.\ D {\bf 88} (2013) no.11,  112006
doi:10.1103/PhysRevD.88.112006
[arXiv:1310.3675 [hep-ex]];
M.~Aaboud {\it et al.} [ATLAS Collaboration],
JHEP {\bf 1806} (2018) 022
doi:10.1007/JHEP06(2018)022
[arXiv:1712.02118 [hep-ex]].

\bibitem{Olive:2016xmw}
C.~Patrignani {\it et al.} [Particle Data Group],
Chin.\ Phys.\ C {\bf 40} (2016) no.10,  100001.
doi:10.1088/1674-1137/40/10/100001
\bibitem{thirdgen}
A.~M.~Sirunyan {\it et al.} [CMS Collaboration],
Eur.\ Phys.\ J.\ C {\bf 77} (2017) no.5,  327
doi:10.1140/epjc/s10052-017-4853-2
[arXiv:1612.03877 [hep-ex]].
M.~Aaboud {\it et al.} [ATLAS Collaboration],
JHEP {\bf 1809} (2018) 050
doi:10.1007/JHEP09(2018)050
[arXiv:1805.01649 [hep-ex]].
\bibitem{winoNLSP}
V.~Khachatryan {\it et al.} [CMS Collaboration],
Eur.\ Phys.\ J.\ C {\bf 74} (2014) no.9,  3036
doi:10.1140/epjc/s10052-014-3036-7
[arXiv:1405.7570 [hep-ex]].
A.~M.~Sirunyan {\it et al.} [CMS Collaboration],
JHEP {\bf 1803} (2018) 166
doi:10.1007/JHEP03(2018)166
[arXiv:1709.05406 [hep-ex]].
A.~M.~Sirunyan {\it et al.} [CMS Collaboration],
JHEP {\bf 1803} (2018) 160
doi:10.1007/JHEP03(2018)160
[arXiv:1801.03957 [hep-ex]].
\bibitem{winoNLSPATLAS}
M.~Aaboud {\it et al.} [ATLAS Collaboration],
arXiv:1803.02762 [hep-ex].
M.~Aaboud {\it et al.} [ATLAS Collaboration],
arXiv:1806.02293 [hep-ex].

\bibitem{Chen:1996ap}
C.~H.~Chen, M.~Drees and J.~F.~Gunion,
Phys.\ Rev.\ D {\bf 55} (1997) 330
Erratum: [Phys.\ Rev.\ D {\bf 60} (1999) 039901]
doi:10.1103/PhysRevD.60.039901, 10.1103/PhysRevD.55.330
[hep-ph/9607421].
\bibitem{strumia}
M.~Cirelli, N.~Fornengo and A.~Strumia,
Nucl.\ Phys.\ B {\bf 753} (2006) 178
doi:10.1016/j.nuclphysb.2006.07.012
[hep-ph/0512090].
\bibitem{ejc1}
E.~J.~Chun,
JHEP {\bf 0912} (2009) 055
doi:10.1088/1126-6708/2009/12/055
[arXiv:0909.3408 [hep-ph]].

\bibitem{pythia}
T.~Sjostrand, L.~Lonnblad and S.~Mrenna,
[arXiv:hep-ph/0108264].

\bibitem{sarah1}
F.~Staub,
Comput.\ Phys.\ Commun.\  {\bf 181} (2010) 1077
[arXiv:0909.2863 [hep-ph]].

\bibitem{sarah2}
F.~Staub, T.~Ohl, W.~Porod and C.~Speckner,
Comput.\ Phys.\ Commun.\  {\bf 183} (2012) 2165
[arXiv:1109.5147 [hep-ph]].


\bibitem{calchep}
A.~Pukhov,
hep-ph/0412191.
A.~Belyaev, N.~D.~Christensen and A.~Pukhov,
Comput.\ Phys.\ Commun.\  {\bf 184} (2013) 1729
[arXiv:1207.6082 [hep-ph]].



\bibitem{lhe}
See "https://pythia6.hepforge.org/examples/"

\bibitem{fastjet}
M.~Cacciari, G.~P.~Salam and G.~Soyez,
Eur.\ Phys.\ J.\ C {\bf 72} (2012) 1896
[arXiv:1111.6097 [hep-ph]].



\bibitem{ejc2}
P.~Bandyopadhyay and E.~J.~Chun,
JHEP {\bf 1011} (2010) 006
doi:10.1007/JHEP11(2010)006
[arXiv:1007.2281 [hep-ph]].
\bibitem{spheno}
W.~Porod and F.~Staub,
Comput.\ Phys.\ Commun.\  {\bf 183} (2012) 2458
doi:10.1016/j.cpc.2012.05.021
[arXiv:1104.1573 [hep-ph]].


\bibitem{6teq6l}
H.~L.~Lai {\it et al.}  [CTEQ Collaboration],
Eur.\ Phys.\ J.\ C {\bf 12}, 375 (2000)
[arXiv:hep-ph/9903282].
J.~Pumplin, D.~R.~Stump, J.~Huston, H.~L.~Lai, P.~Nadolsky and W.~K.~Tung,
JHEP {\bf 0207}, 012 (2002)
[arXiv:hep-ph/0201195].

\bibitem{crosswg}
V.~Khachatryan {\it et al.} [CMS Collaboration],
Phys.\ Rev.\ D {\bf 90} (2014) no.9,  092007
doi:10.1103/PhysRevD.90.092007
[arXiv:1409.3168 [hep-ex]].

\url{https://twiki.cern.ch/twiki/bin/view/LHCPhysics/SUSYCrossSections#Cross_sections_for_various_S_AN3}


 \bibitem{pbad}
P.~Bandyopadhyay, B.~Bhattacherjee and A.~Datta,
JHEP {\bf 1003} (2010) 048
doi:10.1007/JHEP03(2010)048
[arXiv:0909.3108 [hep-ph]].
\bibitem{tao}
G.~F.~Giudice, T.~Han, K.~Wang and L.~T.~Wang,
Phys.\ Rev.\ D {\bf 81} (2010) 115011
doi:10.1103/PhysRevD.81.115011
[arXiv:1004.4902 [hep-ph]].
\bibitem{btag}
I.~R.~Tomalin [CMS Collaboration],
J.\ Phys.\ Conf.\ Ser.\  {\bf 110} (2008) 092033.

\bibitem{seesaw}
Y.~Cai, T.~Han, T.~Li and R.~Ruiz,
Front.\ in Phys.\  {\bf 6} (2018) 40
doi:10.3389/fphy.2018.00040
[arXiv:1711.02180 [hep-ph]].
R.~Franceschini, T.~Hambye and A.~Strumia,
Phys.\ Rev.\ D {\bf 78} (2008) 033002
doi:10.1103/PhysRevD.78.033002
[arXiv:0805.1613 [hep-ph]].


\bibitem{dispjets}
CMS Collaboration [CMS Collaboration],
CMS-PAS-EXO-16-003.

\bibitem{unusig}
P.~Bandyopadhyay, P.~Ghosh and S.~Roy,
Phys.\ Rev.\ D {\bf 84} (2011) 115022
doi:10.1103/PhysRevD.84.115022
[arXiv:1012.5762 [hep-ph]].

\bibitem{RHNs}
P.~Bandyopadhyay, E.~J.~Chun and J.~C.~Park,
JHEP {\bf 1106} (2011) 129
doi:10.1007/JHEP06(2011)129
[arXiv:1105.1652 [hep-ph]].
P.~Bandyopadhyay,
JHEP {\bf 1709} (2017) 052
doi:10.1007/JHEP09(2017)052
[arXiv:1511.03842 [hep-ph]].
\end{thebibliography}
\end{document}